\documentclass[12pt,halfline,a4paper]{ouparticle}

\usepackage{graphicx}
\usepackage{amssymb}
\usepackage{caption}
\usepackage{url}
\usepackage{amsmath}
\usepackage{float}
\usepackage{color}
\usepackage{lineno}
\usepackage{ulem}

\newcommand{\beginsupplement}{%
        \setcounter{table}{0}
        \renewcommand{\thetable}{S\arabic{table}}%
        \setcounter{figure}{0}
        \renewcommand{\thefigure}{S\arabic{figure}}%
     }

\graphicspath{{./Images/}}

\begin{document}


\title{Diving into a simple anguilliform swimmer's sensitivity}

\author{
\name{Nicholas A. Battista}
\address{Dept. of Mathematics and Statistics, 2000 Pennington Road, The College of New Jersey, Ewing Township, NJ 08628, USA}
\email{battistn@tcnj.edu}
} 



\abstract{
Computational models of aquatic locomotion range from individual modest simple swimmers in $2D$ to sophisticated $3D$ individual swimmers to complex multi-swimmer models that attempt to parse collective behavioral dynamics. Each of these models contain a multitude of model input parameters to which the model outputs are inherently dependent, i.e., various swimming performance metrics. In this work, the swimming performance's sensitivity to parameters is investigated for an idealized, simple anguilliform swimming model in $2D$. The swimmer considered here propagates forward by dynamically varying its body curvature, similar to motion of a \textit{C. elegan}. The parameter sensitivities were explored with respect to the fluid scale (Reynolds number, $Re$), stroke (undulation) frequency, as well as a kinematic parameter controlling the velocity and acceleration of each upstroke and downstroke. In total, 5000 fluid-structure interaction simulations were performed, each with a unique parameter combination selected via a Sobol sequence. Thus, global sensitivity analysis was performed using Sobol sensitivity analysis. Results indicate that the swimmer's performance is most sensitive to variation in its stroke frequency. Trends in swimming performance were discovered by projecting the performance metrics from the overall $3D$ parameter space onto particular $2D$ subspaces and Pareto-like optimal fronts were identified for swimming efficiency. This work is a natural extension of the parameter explorations of the same model from \cite{Battista:ICB2020}.

}


\date{\today}

\keywords{aquatic locomotion; anguilliform motion; fluid-structure interaction; immersed boundary method; sensitivity analysis; Sobol sensitivity}

\maketitle

%
%

%
%

%
%

\section{Introduction}
\label{sec:intro}

Mathematical models inherently depend on parameter estimation and selection. In the study of disease \cite{Wu:2013} or drug epidemics \cite{BattistaBMB:2019}, transmission and recovery rates are estimated based on available epidemiological data. Similarly, scientists investigating cellular processes, immunology, or oncology may attempt to fine tune model parameters based off experimental data available to investigate novel therapies \cite{Lodhi:2011,Barish:2017,Eriksson:2019}. Moreover, parameter explorations have provided insight into a number of organismal systems, including odor-capture by crabs \cite{Waldrop:2018} or tubular heart pumping \cite{Waldrop:2020}. A chief focus in each of these studies involved not only parameter exploration within a particular parameter space, but the \textit{sensitivity} of model's output to the model's input parameters. A system is \textit{sensitive} to an input parameter, if slight variations in the parameter result in significant changes in the system's output. Global sensitivity analyses quantify the impact of parameter uncertainty (inputs) on the overall model prediction uncertainty (outputs) in a holistic fashion \cite{Saltelli:2002}. These analyses attempt to determine which parameters, or their combinations, are most sensitive for the system output across a given parameter space.



In organismal biology contexts, uncertain parameters, or fluctuations in parameters, could be viewed slightly differently than those in some population epidemiology or cellular contexts. Rather than only focusing on fine tuning model parameters to match experimental data, variations in parameters could be interpreted as biodiversity, even extending towards evolutionary contexts by exploring robustness of a performance metric to parameter sensitivities \cite{Anderson:2015,Munoz:2017,Waldrop:2020}.  In particular, biodiversity has highlighted elegant ``many-to-one mapping" solutions, where different morphologies (form) can lead to similar performance (function) \cite{Wainwright:2005,Wainwright:2007}. However, not all combinations of traits produce similar performance. To that extent, slight variations may to lead non-linear consequences in functional performance \cite{Arnold:2003}. Upon exploring regions of higher performance within performance landscapes, global sensitivity analysis can determine what variation (or combinations thereof) induce the greatest changes in performance. Thus one is able to rank the importance of specific traits for maintaining a particular performance level (within a specific region on the landscape) while assessing its robustness. Furthermore, quantifying such sensitivities could further provide insight into the evolution rate of a given mechanical system \cite{Munoz:2018,Munoz:2019}.

In the spirit of collaboration across disciplines from the 2020 SICB Symposium \textit{Melding Modeling and Morphology: integrating approaches to understand the evolution of form and function}, this paper highlights a formal mathematical approach for assessing the sensitivity of a model's output due its input parameters. This paper explores the global sensitivity analysis of a mathematical model of a biological system - an idealized swimmer that resembles an aquatic nematode. The selection of an fluid-structure interaction model was two fold: (1) to illustrate that such global sensitivity analysis can be performed on a complex model and (2) to highlight the challenges (necessary computational resources) to perform such an analysis on a computational fluid dynamics model. The swimmer itself, in resembling a nematode, propagates itself forward through its fluid environment by varying its body curvature, i.e., bending \cite{Gray:1964,Ghosh:2008,Majmudar:2012,Luersen:2014}, see Figure \ref{fig:Methods_Geo_Curvature}.

Previous experimental work directly focused on \textit{C. elegans} has highlighted that variations in bending frequency impact locomotion performance, both speed and efficiency, at low Reynolds numbers ($\mathrm{Re}<1$) \cite{Luersen:2014}. The Reynolds number is a dimensionless number that is given as a combination of four parameters: two parameters describing the physical properties of the fluid - its density and viscosity, $\rho$ and $\mu$, respectively, and two system parameters - a characteristic length scale, $L$ and characteristic velocity scale, $V$. Mathematically, it is given as 
\begin{equation}
    \label{eq:Re-intro} Re = \frac{\rho LV}{\mu}.
\end{equation}
Low Re corresponds to situations in which viscous forces are much greater than inertial forces. Many computational models of nematodes have only considered the low Reynolds limit, i.e., the Stokes flow Regime ($\mathrm{Re}=0$) \cite{Berman:2013,Gutierrez:2014,Montenegro:2016}. However, anguilliform swimming gaits are observed in other organisms, such as fly larva, leeches, eels, or lamprey \cite{Taylor:1952,Jordan:1998,Tytell:2004,Hamlet:2015}. These organisms are much larger and live at intermediate Re ($\mathrm{Re}\sim10s$ or $100s$) or higher Re ($\mathrm{Re}\gtrsim1000$) ranges. A recent computational study sought out to explore the effectiveness of this anguilliform gait across a variety of fluid scales, $f$, and intrinsic curvature dynamics \cite{Battista:ICB2020}.

This work is a continuation of the work in \cite{Battista:ICB2020}, which explored the same idealized, simple swimmer's anguilliform gait and swimming performance through thorough extensive parameter subspace explorations. The parameter space considered was composed a fluid scaling parameter, given by an input Reynolds number ($\mathrm{Re}_{in}$), the stroke (undulation) frequency ($f$), and a kinematic control parameter ($p$). The kinematic parameter controls the velocity and acceleration of the swimmer's body curvature as it undulates between a concave up and concave down state. That is, it controls each stroke's acceleration from rest in its current curvature (concavity) state to its maximal velocity and back to rest in its next curvature (concavity) state. Note that previous experimental work suggests that a nematode's undulatory amplitude does not change when places in increasing viscosity environments \cite{Korta:2007}, i.e., decreasing Re settings; however, this work will show that does not appear be the case as Re increases. 

While \cite{Battista:ICB2020} performed extensive parameter explorations using the same anguilliform swimmer model, specific 2D slices through the overall 3D parameter space were chosen in which to investigate swimming performance. The selected 2D slices were always perpendicular to other parameter in the 3D space. For example, as the entire parameter space considered was $\mathrm{Re}_{in}\times f\times p=[0.3,4500]\times[1,2.5]\times[0.075,0.425]$, one 2D slice could be all (Re$,f)$ combinations for $p=0.25$. Note that the range of $\mathrm{Re}_{in}$ given above is based off the definition of input Reynolds number provided in this manuscript, later in Eq. \ref{eq:Re}). Although this systematic approach allows one to identify a parameter subspace to which results in desired swimming performance, i.e., desired threshold forward swimming speeds or efficiencies, it has a few significant drawbacks. First, it is unclear whether the resulting performance data could be interpolated with any accuracy between parallel 2D slices due to nonlinear relationships that may exist. Second, it is restrictive in that it does not allow one to perform a formal global sensitivity analyses on 3D parameter space, due to the manner in which the 3D space was sampled. Thus you cannot determine which parameter (or combinations thereof) most significantly affect the model's output across the input parameter ranges considered.

The anguilliform swimmer model explored here has substantial forward swimming speeds when $\mathrm{Re}_{in}\gtrsim450$ \cite{Battista:ICB2020}. Therefore, with the goal of performing global sensitivity analyses, the parameter subspace selected to perform the analyses was $\mathrm{Re}\in[450,2200], f\in[1,3]\ \mathrm{Hz}$, and $p\in[0.05,0.45].$ In the remainder of this paper, Sobol sensitivity analysis \cite{Sobol:2001} is performed on 5 different swimming performance metrics on this specified subspace described above. By comparing 5 different output metrics sensitivity to parameters it is possible to determine how different performance metrics may have differing parameter sensitivities, both in determining which parameter it is most sensitive to as well as its relative importance (magnitude). Discernible patterns are also uncovered in swimming performance by projecting the $3D$ parameter space sampled onto specific $2D$ subspaces. Furthermore, Pareto-like optimal fronts in swimming efficiency are identified across various combinations of the input parameters \cite{Eloy:2013, Verma:2017,Smits:2019}. Lastly, the parameter sensitivity suggested by these Pareto-like fronts is consistent with that of the formal Sobol sensitivity analysis. Both Sobol sensitivity analysis and Pareto-like optimal fronts will be described in further detail in Sections \ref{sec:methods} and \ref{results}, respectively.

\section{Methods}
\label{sec:methods}

\begin{figure}[H]
    \centering
    \includegraphics[width=0.90\textwidth]{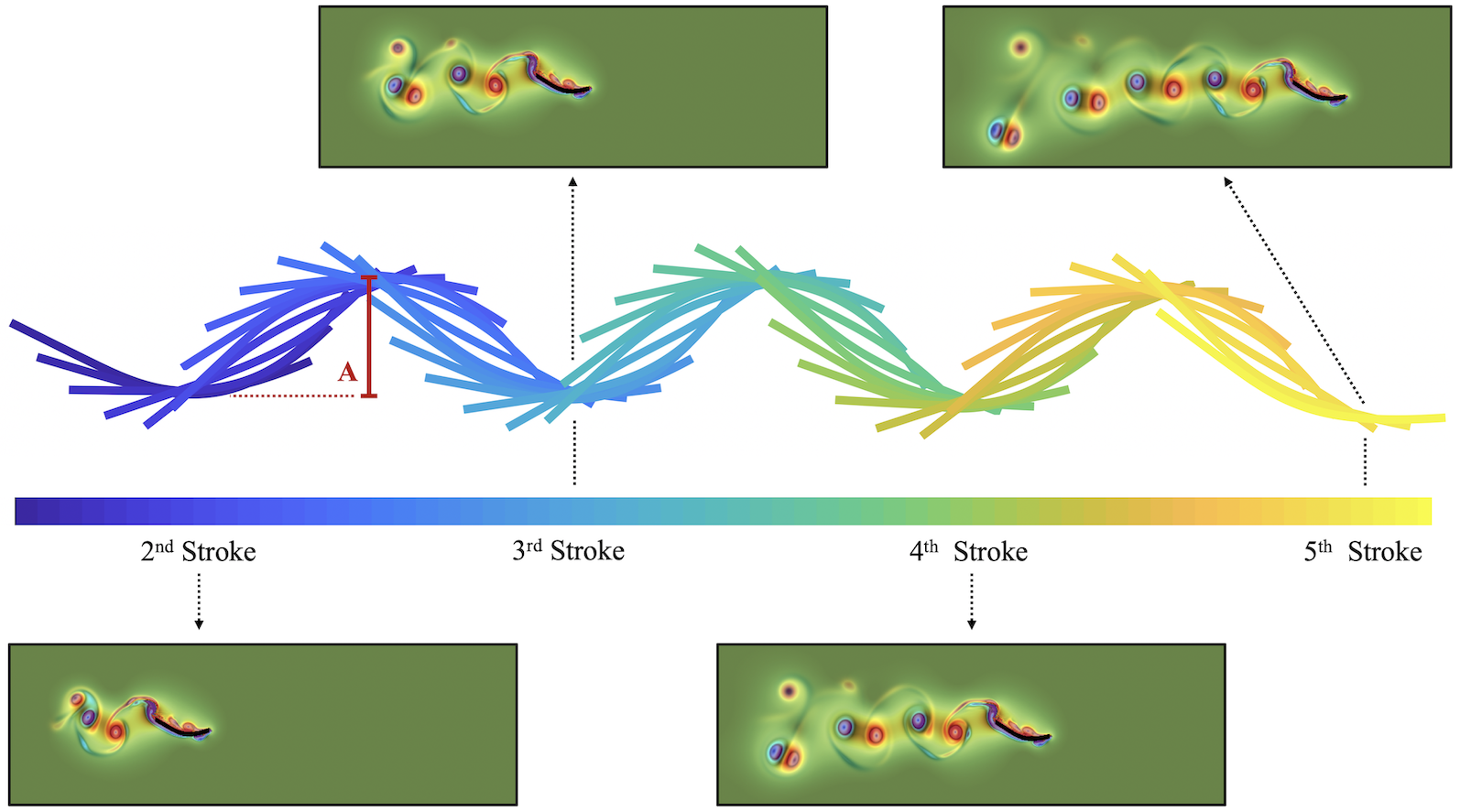}
    \caption{An illustration of how the swimmer changes its preferred curvature to propagate forward during the $2^{nd}$ through $5^{th}$ stroke cycles. Snapshots of the swimmer and the associated fluid vorticity (colormap) are given at the beginning of each successive stroke cycle. The definition of the peak-to-peak amplitude, $A$, is given. This amplitude is an output of the model.}
    \label{fig:Methods_Geo_Curvature}
\end{figure}

In this work, I continue the investigation of the idealized, simple anguilliform swimming model first presented in \cite{BattistaIB2d:2018} and further explored in \cite{Battista:2020,Battista:ICB2020}. To propagate forward, i.e., swim, this swimmer changes the curvature along its $1D$ body, which resembles aquatic nematode locomotion \cite{Jung:2008,Majmudar:2012}, see Figure \ref{fig:Methods_Geo_Curvature}. Its total stroke (undulation) cycle is compromised of both an upstroke and downstroke, which end when the swimmer's body is either concave down and concave up, respectively. As the swimmer propagates forward, it leaves behind a vortex wake. This observation suggests that hydrodynamic force generation by dynamically changing curvature is sufficient for locomotion \cite{Rayner:1995}. The swimmer given in Figure \ref{fig:Methods_Geo_Curvature} corresponds to the case of $(\mathrm{Re}_{in},f,p)=(1011.75, 1.6875, 0.4125)$. As these parameters are varied, the swimmer's performance may substantially change, as can be seen in Figures \ref{fig:Colormap_Speed_COT}, \ref{fig:Colormap_St_Deff}, and \ref{fig:Results_COT_vs_Speed} (and as shown in \cite{Battista:ICB2020}).

I used the open-source fluid-structure interaction software, \textit{IB2d} \cite{Battista:2015,BattistaIB2d:2017,BattistaIB2d:2018}, based on the immersed boundary method (IB) \cite{Peskin:2002} to perform the computational fluid dynamics simulations. This framework allows one to study this particular anguilliform swimming mode in an incompressible, viscous fluid, thus allowing for studies across numerous fluid scales i.e., Reynolds numbers. Note that this anguilliform model is one of the built-in examples contained in the software offered to the scientific community. It can be found at \url{github.com/nickabattista/IB2d} in the sub-directory: 
\begin{center}
\texttt{IB2d$/$matIB2d$/$Examples$/$Examples$\_$Education$/$Interpolation$/$Swimmer}.
\end{center}

A background introduction to IB and the swimmer's implementation are provided in the Supplemental Materials; greater detail regarding the mathematical framework, formulation, and implementation can be found in \cite{BattistaIB2d:2018,Battista:2020,Battista:ICB2020}. The computational grid, temporal, and material property parameters are identical to those used in \cite{Battista:ICB2020}; they are also given in Table \ref{table:num_param} in the Supplemental Materials. The model parameters varied in this study are an \textit{input} Reynolds Number, $\mathrm{Re}_{in}$, stroke (undulation) frequency, $f$, and a kinematic control parameter, $p$. The kinematic parameter, $p$, helps govern the kinematic profile of each stroke. Although, each upstroke (or downstroke) will always perform an undulatory pattern, varying $p$ is akin to changing the acceleration and velocity through stroke pattern itself. That is, it controls how quickly each stroke accelerates from rest in the starting curvature (concavity) state, to its maximal velocity, and back to rest in the other curvature (concavity) state. 

As many models of anguilliform motion assume complex sinusoidal curvature kinematics \cite{Kern:2006,VanRees:2015}, the curvature interpolation scheme used here provides functionality to subtly change properties of the stroke, i.e., the maximum velocity and accelerations of the changing body curvature during each upstroke or downstroke. Given two (or more) curvature (geometric) states, one could design a custom interpolation function(s) based on an animal's kinematic data. This could be done by tracking points along the moving body and in addition to recording the position over time, using such data to approximate the points' associated velocity and acceleration between different positions. The curvature interpolant here used a cubic spline interpolant \cite{Battista:2020,Battista:ICB2020}, which allowed for one free parameter ($p$) that could be varied to change the body curvature's maximal velocity and acceleration. Increasing $p$ would continue to deviate away from pure sinusoidal behavior, see Figure \ref{fig:interpolation} in the Supplemental Materials. Moreover, this interpolation frame also allowed me to enforce that no instantaneous accelerations occur along the swimmer's body, unlike trivial sinusoidal curvature interpolants, see Figure \ref{fig:interpolation}c. Introducing a higher-order interpolation function would give rise to more free parameters which would allow one to more closely replicate the exact kinematics of a moving body.

In other words, in this study I varied one parameter that encompasses size variations (the input Reynolds number, $\mathrm{Re}_{in}$), another parameter that details how often swimming movement patterns are performed (the frequency, $f$), and a third that changes the dynamics within the kinematics of how each stroke is performed (kinematic parameter, $p$). Note that the ranges of the model parameters were slightly different than those in \cite{Battista:ICB2020}. Here a $3D$ parameter subspace was selected based on the previous explorations in which there was significant forward swimming speed: $\mathrm{Re}_{in}\times f\times p = [450,2200]\times[1,3]\ \mathrm{Hz} \times[0.05,0.45]$. Note that the range of $\mathrm{Re}_{in}$ here is identical to the range of $\mathrm{Re}_{in}\in[150,750]$ in \cite{Battista:ICB2020}, based on that work's definition of Re. The range of $f$ was selected to center about the undulation frequency that \textit{C. elegans} display while swimming in water, roughly $2$ Hz \cite{Luersen:2014,Backholm:2015}. Although, in wet granular media \textit{C. elegans} swim with frequencies around $\sim 1$ Hz as well as exhibit faster swimming behavior \cite{Jung:2008,Jung:2010}. On the other hand, \textit{C. elegans} undergoing dietary restrictions have been observed undulating at higher frequencies of $\sim 3\ \mathrm{Hz}$, also swam faster than the standard control case \cite{Luersen:2014}. Other anguilliform swimmers such as eels and lampreys have been observed swimming at higher frequencies. For example, eels have been observed to steadily swim with an undulation frequency of $\sim 3$ Hz \cite{Tytell:2004a}. Thus, the range of frequencies was chosen to be between [1,3] Hz. The values of the kinematic parameter $p$ were chosen to span the range in which lead to a symmetric interpolation function, see Figure \ref{fig:interpolation} in the Supplemental Materials.

\begin{figure}[H]
    \centering
    \includegraphics[width=0.90\textwidth]{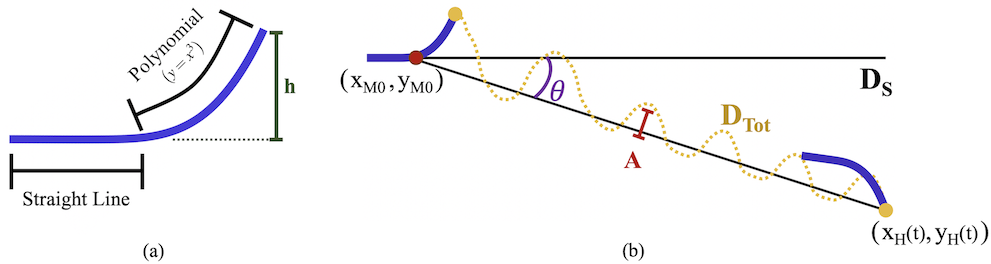}
    \caption{(a) The swimmer's geometry is composed of a straight line segment and a curved portion, given by a cubic polynomial. The swimmer's height, $h$, is the height of the curved portion off the straight segment. It remained uniform across all simulations. (b) The definitions for the angular trajectory off the horizontal ($\theta$), as measured off the swimmer's head and a fixed point along the horizontal (last point along the straight portion of the swimmer's body), the horizontal distance swam ($D_S$), and the total linear distance moved by the swimmer ($D_{Tot}$). The swimmer in (b) corresponds to the case of ($\textit{Re}_{in},f,p)=(270,1.25,0.25)$. Figure adapted from \cite{Battista:ICB2020}.}
    \label{fig:Methods_Geo_Curvature-2}
\end{figure}

To achieve the desired input Reynolds number, $\mathrm{Re}_{in}$, first a tuple was selected, i.e., $\mathrm{Re}_{in},f,p)$, followed by finding the appropriate dynamic viscosity, $\mu$, to give the appropriate $\mathrm{Re}_{in}$, i.e.,

\begin{equation}
    \label{eq:Re} \mathrm{Re}_{in} = \frac{\rho L (fL)}{\mu}, 
\end{equation}
where $L$ is the swimmer's bodylength (characteristic length) and the product $fL$ gives an input frequency-based velocity scale. Note that in anguilliform studies it is common to use the product of $f$ and a peak-to-peak undulation amplitude, $A$, as the characteristic velocity in the Reynolds number calculation. However, since undulation amplitude is an output of the model and cannot be known \textit{a priori}, I elected to use a characteristic velocity seen in fish literature as an input velocity scale, $fL$ \cite{Cui:2017,Dai:2018}. Illustrations of the peak-to-peak amplitude, $A$ are given in Figures \ref{fig:Methods_Geo_Curvature} and \ref{fig:Methods_Geo_Curvature-2}. Plots of the input Reynolds number ($\mathrm{Re}_{in}$ vs output Reynolds number ($\mathrm{Re}_{out}$) organized by parameters are given in Figure \ref{fig:input_Re_vs_output_Re} in the Supplemental Materials. The output Reynolds number is defined as $Re_{out}=\rho\cdot L\cdot fA/\mu$, whose frequency based velocity scale is given as $fA$. 

Six swimming performance metrics were computed for each simulation. The motivation behind calculating 6 metrics was to identify whether different performance metrics were most sensitive to different input parameters. Rather than report the dimensional swimming speed, $V_{dim}$, for each simulation, all swimming speeds are provided in non-dimensional form, given by the inverse of the Strouhal number. The Strouhal number is defined as
\begin{equation}
    \label{eq:Strouhal} St = \frac{fA}{V_{dim}}.
\end{equation}
where $A$ is the peak-to-peak undulation amplitude, see Figures \ref{fig:Methods_Geo_Curvature} and \ref{fig:Methods_Geo_Curvature-2}. Note that $A$ is an output of the model itself, whose changing values due to variations input parameter changes was also analyzed.

The Strouhal number has been previously to assess efficiency swimming \cite{Taylor:2003,Eloy:2012}. It compares the wavespeed to the average speed of the swimmer. I will also compute three other efficiency metrics: a power-based cost of transport \cite{Bale:2014,Hamlet:2015}, $COT$, a distance effectiveness ratio, $d_{eff}$, and an angular trajectory metric, $\theta$. The $COT$ provides the energetic cost per unit speed of the swimmer, $d_{eff}$ compares the forward distance swam to the total linear distance swam overall, and $\theta$ indicates how horizontal the overall swimming movement stayed. All of these quantities are provided in non-dimensional units, i.e.,
\begin{align}
    \label{eq:COT} COT &= \frac{1}{N}\ \frac{1}{V_{dim}}\ \frac{1}{\rho (fA)^2 L^2} \displaystyle\sum_{j=1}^N |F_j||U_j| \\
    \label{eq:DistEff} d_{eff} &= \frac{D_S}{D_{Tot}} \\
    \label{eq:Angle} \theta &= \frac{1}{N} \displaystyle\sum_{j=1}^N \tan^{-1}\left( \frac{Y_{H}(t)-Y_{M0}}{X_{H}(t)-x_{M0}}\right)
\end{align}
where $D_{S}$ and $D_{Tot}$ are the forward distance swam (horizontal distance) and total linear distance moved by the swimmer during a specific period of time encompassing $N$ time-points. $F_j$ and $U_{r_{j}}$ are the applied force and tangential body velocity of the swimmer, at the $j^{th}$ time step, respectively. The applied force was computed as the force perpendicular to the original direction of motion of the swimmer. $X_{H}(t)$ and $Y_{H}(t)$ are the time-dependent Lagrangian position of the swimmer's head and $(X_{M0},Y_{M0})$ is the initial position of the last point along the straight section of the swimmer's body (see Figure \ref{fig:Methods_Geo_Curvature-2}). The non-dimensional $COT$ is similar the energy-consumption coefficient in \cite{Bale:2014}. Lastly to complement $d_{eff}$, the average angle off the horizontal, $\theta$, was calculated. This allows one to discern parameter combinations that result in non-horizontal swimming trajectories.

%

To assess this swimming model's overall sensitivity to its parameters, I used Saltelli's extension of the Sobol sequence \cite{Saltelli:2002,Saltelli:2010} to form a discrete subset of tuples of the overall parameter space in which to simulate the swimmer, i.e., a total of N tuples were selected via the Sobol sequence process, $\{(\mathrm{Re}_{in},f,p)_j\}_{j=1}^N$, to run the model for. Once the simulations were performed, they were analyzed to produce a peak-to-peak undulation amplitude $A$ and a Strouhal number, $St$, and thereby a time-averaged forward swimming speed ($1/St$), $d_{eff}$, $COT$, and $\theta$. Sobol sensitivity analysis was then performed for each swimming performance metric. Sobol sensitivity is a variance-based sensitivity analysis that can provide \textit{global} sensitivity to parameters, rather than only local sensitivity \cite{Link:2018}. By quantifying global sensitivity, one can determine which parameter, when varied within a particular range, results in the most significant changes in the model's output, even with respect to other parameters being varied. Moreover, Sobol analysis is able to efficiently calculate first-order parameter sensitivity indices, i.e., perturbations of one parameter at a time, but also higher-order indices, i.e., those corresponding to perturbations of two or more parameters at a time, and total-order indices, i.e., all combinations of other parameters \cite{Saltelli:2010}. Furthermore, due to this, the importance of higher-order interactions can be inferred by comparing first-order and total-order sensitivity indices. If there are significant differences between these indices, it suggests the presence of higher-order interactions. Higher-order interactions occur when two or more parameters are changed and it causes a greater variation in the output than when varying each of those inputs alone. \textit{Local} sensitivity analyses, on the other hand, have the misfortune of necessitating that only one parameter can be varied at a time, which considerably restricts the parameter space that is able to be explored and analyzed. Under-resolving the input parameter space could easily lead to inaccurate sensitivities, unless the model is linear \cite{Saltelli:2019}. Moreover, higher-order interactions are difficult to accurately parse out of the data using only local sensitivity methods. The model examined here was previously seen to exhibit non-linear behavior \cite{Battista:ICB2020}. Therefore it was deemed necessary to use global approach for sensitivity.

\begin{figure}[H]
    \centering
    \includegraphics[width=0.99\textwidth]{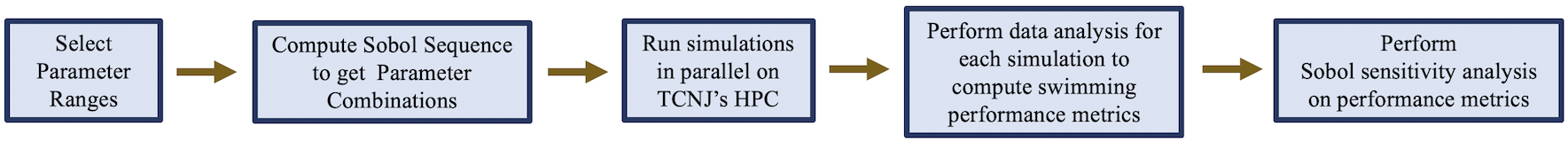}
    \caption{Workflow of the entire simulation and sensitivity process}
    \label{fig:Methods_Workflow}
\end{figure}

The workflow for the entire simulation and sensitivity analysis process is illustrated in Figure \ref{fig:Methods_Workflow}. 

\section{Results}
\label{results}

A Sobol sequence was generated using $N=1000$ and $d=3$ (the dimension of the parameter space) using Saltelli's extension of the Sobol sequence \cite{Saltelli:2002,Saltelli:2010} for a total of $N(d+2)=5000$ different parameter combinations to be simulated in order to find the Sobol sensitivity indices of each performance metric. The simulations were performed on The College of New Jersey's high performance computing cluster \cite{TCNJ:ELSA}. As each simulation took approximately $\sim24$ hours to run, the entire effort of this study took required approximately a total of 120,000 computational hours. 

\begin{figure}[H]
    \centering
    \includegraphics[width=0.90\textwidth]{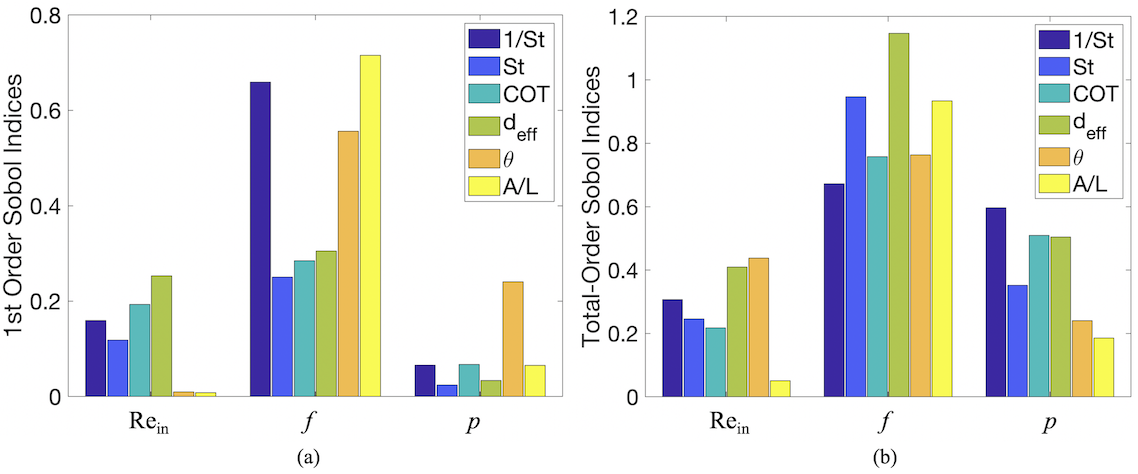}
    \caption{(a) First-order and (b) Total-order Sobol indices of the three varied parameters Re, $f$, and $p$ for 5 swimming performance metrics: non-dimensional swimming speed ($1/St$), Strouhal number ($St$), non-dimensional cost of transport ($COT$), a distance effectiveness ratio ($d_{eff}$), and the average angular trajectory from horizontal ($\theta$).} 
    \label{fig:Sobol_Indices}
\end{figure}

Once the swimming performance metrics were computed, the Sobol sensitivity indices were computed \cite{Sobol:2001,Saltelli:2010}. Figure \ref{fig:Sobol_Indices} provides the indices for both the first-order and total-order parameter interactions. Over all the model parameter ranges considered, the swimming performance metrics quantified here were most sensitive to variations in stroke (undulation) frequency, $f$. Moreover, there appear to exist higher-order interactions, as the first-order indices and total-order indices are not equivalent; however, the same trend exists of the system being most sensitive to $f$. This suggests that the efficiency metrics explored here were most sensitive to the same parameter, that parameter being the frequency, $f$. This was remains true even for the dimensional analogs of swimming speed and cost of transport, see Figure \ref{fig:Sobol_Indices_2} in the Supplemental Materials. Moreover, the degree of sensitivity to each parameter was varied across all the output metrics. Note that while this study only varied $\mathrm{Re}_{in}, f$, and $p$, it is possible that the efficiency metrics could be more sensitive to other parameters not explored here, such as the degree of body curvature, amplitude of each stroke, body thickness, or muscular activation force \cite{Gillis:1996,Padmanabhan:2012,VanRees:2013,Williams:2015}. 

In the restricted $3D$ parameter comprising $\mathrm{Re}_{in}\times f\times p = [450,2200]\times[1,3]\times[0.075,0.425]$, swimming performance appears most sensitive to stroke frequency, $f$. With only the Sobol indices in hand, it is unclear in what particular ways the performance metrics are sensitive to $f$, e.g., if $f$ is increased, it is unclear whether swimming speeds increase or decrease. More so, it isn't clear how to determine a finer parameter subspace where higher swimming performance resides directly from the Sobol sensitivity results. Furthermore, even if a finer subspace was tested, since the ranges of the model parameters would be different in that subspace, it would inherently change the sensitivity indices, and in possible significant ways as well.

\begin{figure}[H]
    \centering
    \includegraphics[width=0.90\textwidth]{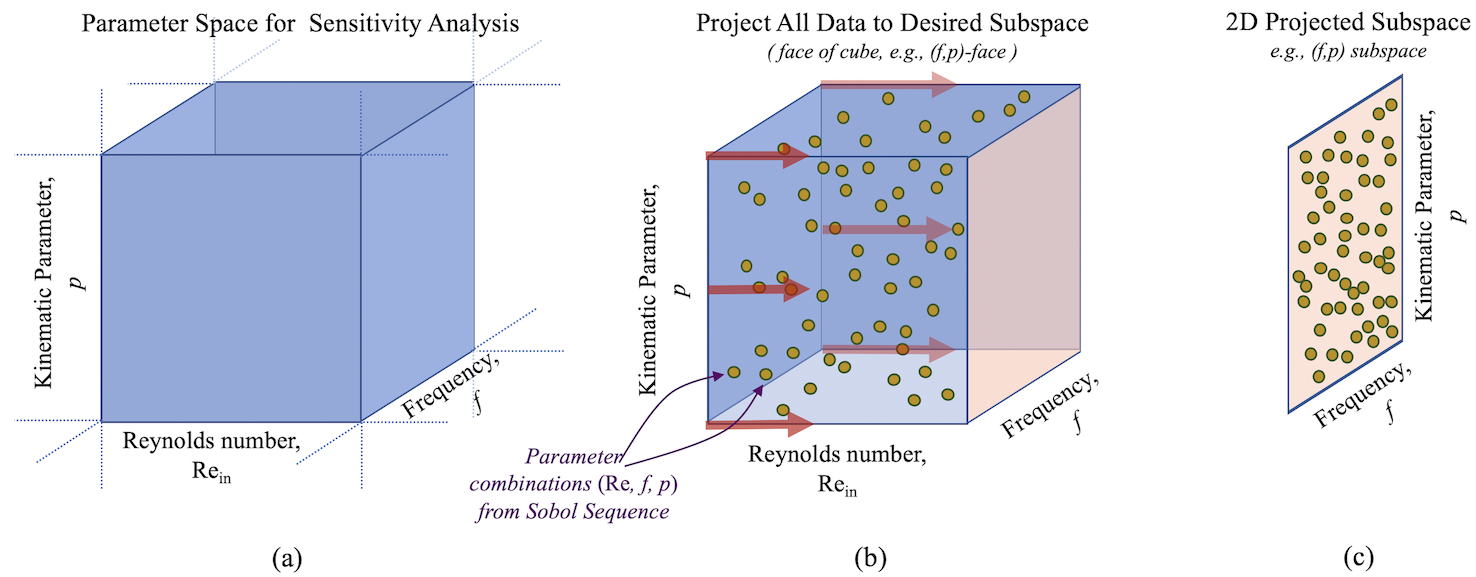}
    \caption{(a) The overall $3D$ parameter space being sampled by Sobol sequences. (b) Illustration of sampling points via Sobol sequences being projected into a $2D$ subspace, here the $(f,p)$ subspace (c) Visualization of where data is projected from a higher dimension space into a $2D$ subspace.}
    \label{fig:Projection}
\end{figure}

From the Sobol sensitivity analysis performed in this paper, two-dimensional subspace colormaps are not able to be produced in the same manner as in \cite{Battista:ICB2020}. Rather than coplanar $2D$ subspaces always existing on a plane perpendicular to one of the axial dimensions, e.g., the $(\mathrm{Re}_{in},f)$-subspace for a particular $p$ value as in \cite{Battista:ICB2020}, the selected $3D$ parameter space was sampled using Sobol sequences, see Figure \ref{fig:Projection}. Thus, similar `flat', co-planar subspaces do not exist in which many of the sampled points lie. Hence the entire parameter space sampled via Sobol sequences was projected into three distinct subspaces: $(\mathrm{Re}_{in},f)$, $(\mathrm{Re}_{in},p)$, and $(f,p)$, as in process illustrated by Figures \ref{fig:Projection}b-c. Having performed 5000 simulations, these manifest themselves in projected subspaces, which appear ``filled", to which may be parsed for general performance trends.

\begin{figure}[H]
    \centering
    \includegraphics[width=0.95\textwidth]{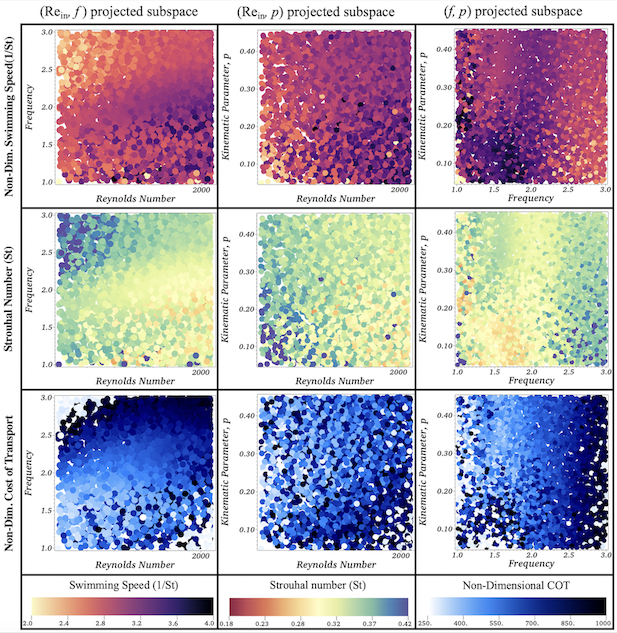}
    \caption{Colormaps corresponding to the non-dimensional forward swimming speeds ($1/St$), Strouhal numbers ($St$), and cost of transports ($COT$) for all the data sampled from Sobol sequences when projected onto either the $(\mathrm{Re}_{in},f)$, $(\mathrm{Re}_{in},p)$, or $(f,p)$ subspaces.}
    \label{fig:Colormap_Speed_COT}
\end{figure}

\begin{figure}[H]
    \centering
    \includegraphics[width=0.95\textwidth]{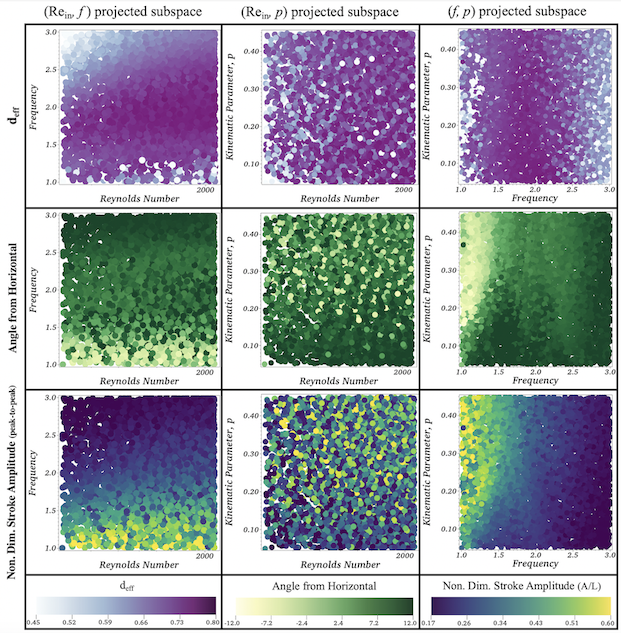}
    \caption{Colormaps corresponding to the distance effectiveness ratio ($d_{eff}$), the angular trajectory off the horizontal ($\theta$), and non-dimensional stroke amplitude ($A/BL$) for all the data sampled from Sobol sequences when projected onto either the $(\mathrm{Re}_{in},f)$, $(\mathrm{Re}_{in},p)$, or $(f,p)$ subspaces.}
    \label{fig:Colormap_St_Deff}
\end{figure}

Figures \ref{fig:Colormap_Speed_COT} and \ref{fig:Colormap_St_Deff} provide the projected data onto each respective subspace: $(\mathrm{Re}_{in},f)$, $(\mathrm{Re}_{in},p)$, or $(f,p)$. From a quick qualitative glance, patterns emerge within regions of the $(\mathrm{Re}_{in},f)$ and $(f,p)$ subspaces for all performance metrics, while in the $(\mathrm{Re}_{in},p)$ subspace, the data looks appears like random noise. For example, the cost of transport, $d_{eff}$, $\theta$, or $A$ panels corresponding to the $(\mathrm{Re}_{in},p)$ subspace do not show much, if any, immediate discernible qualitative pattern(s). However, it does appear that lower $f$ corresponds to more noisy data within almost every subspace. Furthermore, note that the subspaces in which patterns emerged had $f$ as one of its component axes. Moreover, the patterns and trends emerge within the performance data in the direction of $f$-axis in both of these subspaces, i.e., the $(\mathrm{Re}_{in},f)$ and $(f,p)$ subspaces. While slight variations do occur in the other component's direction, either $\mathrm{Re}_{in}$ or $p$, in these two subspaces, respectively, qualitatively the most substantial change occurs in the direction of $f$. This agrees with the Sobol sensitivity analysis above, see Figure \ref{fig:Sobol_Indices}, which indicated that swimming performance was most sensitive to the the stroke (undulation) frequency, $f$. Although, as these plots suggest, $f$ is not the only significantly contributing parameter to the model performance output.

The patterns that emerged in Figures \ref{fig:Colormap_Speed_COT} and \ref{fig:Colormap_St_Deff} show some of the same general trends as those seen in \cite{Battista:ICB2020}. For example, the high swimming speeds arose from high $\mathrm{Re}_{in}$ and lower $f$; however, here this parameter region also corresponded to lower $COT$. Interestingly, the minimal value regions for $COT$ fell either for high $\mathrm{Re}_{in}$ and low $f$ or low $\mathrm{Re}_{in}$ and high $f$. Although, low $\mathrm{Re}_{in}$ and high $f$ correspond to a region of higher $St$, where $St$ leaves the efficient biological range of $0.2<St<0.4$ \cite{Taylor:2003}. High values of $d_{eff}$ in the $(\mathrm{Re}_{in},f)$ subspace also corresponded to lower values in $COT$ and moderately high swimming speeds, in which the swimmers swam with emergent peak-to-peak stroke amplitudes of approximately 1/3 of their bodylength. These occurred around $f\sim2\ \mathrm{Hz}$. \textit{C. elegans} have been observed swimming at roughly $f\sim 2\ \mathrm{Hz}$ with peak-to-peak amplitudes that were roughly 25\% of their bodylength, but at $Re\approx0.5$ \cite{Jung:2010}. It appeared overall that low frequency swimming resulted in higher peak-to-peak stroke amplitudes and downward swimming trajectories. Moreover, the performance quantities show non-linear dependence on $f$ and $p$, as inferred from the $(f,p)$ panels. Note that the trends seen in the $(\mathrm{Re}_{in},f)$ and $(\mathrm{Re}_{in},p)$ projected subspaces  are consistent with those across the projected subspaces involving the output Reynolds number, $\mathrm{Re}_{out}$, i.e., $(\mathrm{Re}_{out},f)$ and $(\mathrm{Re}_{out},p)$. This data is provided in Figures \ref{fig:Colormap_RealRe_Freq} and \ref{fig:Colormap_RealRe_P} in the Supplemental Materials. 


\begin{figure}[H]
    \centering
    \includegraphics[width=0.825\textwidth]{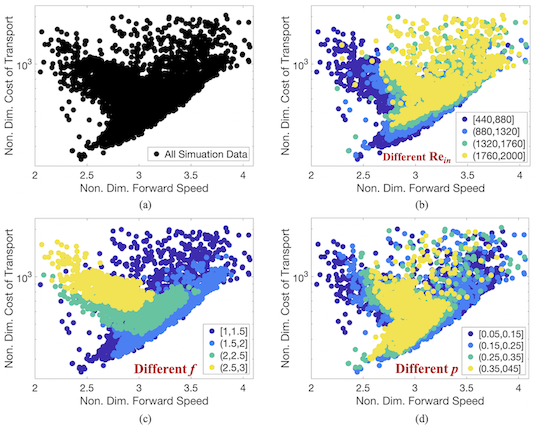}
    \caption{The non-dimensional $COT$ and forward swimming speeds ($1/St$) plotted against each other for (a) every simulation performed, as well as indicating clusters corresponding to differing ranges of the input parameters: (b) different $\mathrm{Re}_{in}$ (c) different $f$, and (d) different $p$.}
    \label{fig:Results_COT_vs_Speed}
\end{figure}

\begin{figure}[H]
    \centering
    \includegraphics[width=0.95\textwidth]{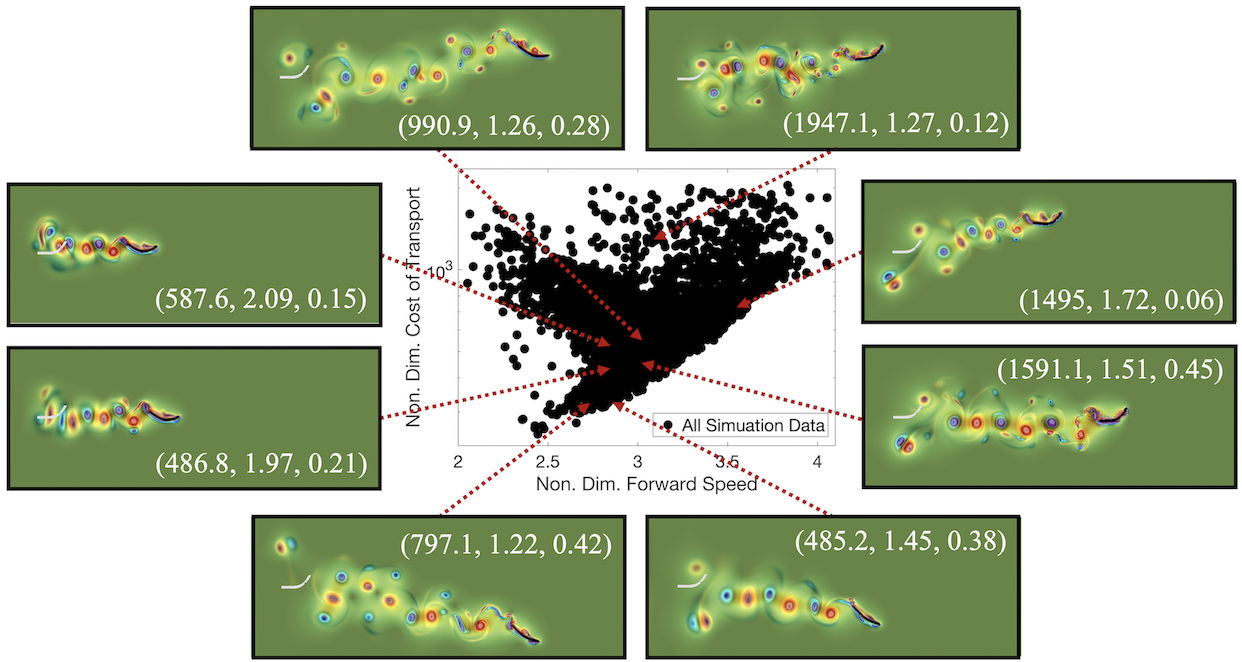}
    \caption{Numerous swimmer's vortex wake and position after its $5^{th}$ full stroke cycle as well as indicating where they fall onto the non-dimensional cost of transport vs. swimming speed performance space. The colormap illustrates vorticity and the greyed out swimmer shows the starting position of the swimmer in each case. Different parameter combinations $(\mathrm{Re}_{in},f,p)$ lead to different swimming behavior, as indicated by its placement within the performance space and vortex wake left behind.}
    \label{fig:Pareto_Vorticity_NonDim}
\end{figure}

Lastly, a Pareto-like optimal front is identified when plotting the non-dimensional cost of transport ($COT$) against the non-dimensional swimming speed ($1/St$) \cite{Eloy:2013,Verma:2017,Schuech:2019,Smits:2019,Tokic:2019}, see Figure \ref{fig:Results_COT_vs_Speed}a. This is referred to as \textit{Pareto-like} simply because minimal $COT$ is desired for maximal $1/St$, thus both metrics are not being maximized as is standard in traditional Pareto optimization. The $COT$ appears to increase exponentially (note the logarithmic axis) with increasing swimming speed.  Figures \ref{fig:Results_COT_vs_Speed}b-d, depict where different input parameter ranges lie within the performance landscape across all simulations performed. (b) and (d) illustrate that for a given $\mathrm{Re}_{in}$ or $p$ within the input parameter space that combinations of the other two parameters, either $(f,p)$ or $\mathrm{Re}_{in},f)$, respectively, could result in a swimmer whose performance could almost be anywhere in the entire landscape. For example, given $\mathrm{Re}_{in}=500$, one could create a swimmer whose swimming speed was anywhere between $[2,4]$ by selecting the appropriate $f$ and $p$ combination. On the other hand, distinct clusters emerge in (c) for different $f$. As $f$ increases, both swimming speed decreases while $COT$ increases generally. Undulation frequencies within the range of $[1,2]$ extend the length of the Pareto-like front. However, frequencies within the $[1,1.5]$ range span the entire performance space, while frequencies within $[1.5,2]$ are more restricted towards the Pareto-like front. This further supports that the model output is most sensitive to undulation frequency, $f$, within the aforementioned parameter subspace considered in this study. Where different parameter combination swimmers lie within the performance space is given in Figure \ref{fig:Pareto_Vorticity_NonDim}. This figure also illustrates those swimmer's position and vortex wakes after their $5^{th}$ stroke cycle. There do not appear to be any qualitative patterns among the vortex wakes that suggest higher or lower cost of transport. This performance space data is also given in \textit{dimensional} form of cost of transport (N/kg) and swimming speed (bodylength/s) in Figures \ref{fig:DIM_COT_vs_Speed} and \ref{fig:Pareto_Vorticity_Dim} in the Supplemental Materials. The same swimmers were used in Figures \ref{fig:Pareto_Vorticity_NonDim} and \ref{fig:Pareto_Vorticity_Dim}. A similar trend is observed in the dimensional data, in which distinct clusters form for different $f$, as in Figure \ref{fig:Results_COT_vs_Speed}c, although higher $f$ corresponds to higher dimensional cost of transport. Moreover, there appears to be more clustering for different $\mathrm{Re}_{out}$ than in Figure \ref{fig:Results_COT_vs_Speed}b, where higher $\mathrm{Re}_{out}$ appear to correspond to faster swimming (bodylength/s) and slightly lower cost of transport (N/kg).

\section{Discussion}
\label{discussion}

Previous experimental and computational studies have highlighted the importance of size (scale) \cite{Alben:2012c,Shelton:2014}, stroke/undulation frequency \cite{Bainbridge:1958,Gray:1968,Steinhausen:2005}, bending curvature and stiffness \cite{Shelton:2014,Hoover:2018}, and intrinsic kinematics \cite{VanRees:2015,Hamlet:2018} on swimming performance. In this work, 3 input parameters were varied: the scale ($\mathrm{Re}_{in}$), stoke (undulation) frequency ($f$), and a kinematic control parameter ($p$) for an idealized, simple anguilliform fluid-structure interaction (FSI) model, that resembles a nematode. The input parameter space was composed entirely of a region in which there is substantial forward swimming performance \cite{Battista:ICB2020}. Rather than focusing on the net effect that variations of a single parameter has on swimming performance, the model's global sensitivity to its input parameters was uncovered using Sobol sensitivity analysis.

In order for the Sobol sensitivity indices to converge, a total of 5000 simulations were performed, requiring approximately 120,000 computational hours on the The College of New Jersey's HPC \cite{TCNJ:ELSA}. The results yielded that in the parameter space considered ($\mathrm{Re}_{in}\times f\times p = [450,2200]\times[1,3]\ \mathrm{Hz} \times[0.05,0.45]$) that swimming performance was most sensitive to variations in the stroke (undulation) frequency (Figures \ref{fig:Sobol_Indices} and \ref{fig:Sobol_Indices_2}). 


While it is already well-known that frequency affects forward speed and performance \cite{Bainbridge:1958,Gray:1968,Steinhausen:2005}, this study indicates that varying the frequency (within the parameter space considered) will more significantly affect the resulting model's performance than changes in the other two input parameters and combinations thereof. In the parameter space considered, all the performance metrics considered were most sensitive to frequency; however, the degree of sensitivity to each parameter varied between different metrics. For example, after frequency, the total-order Sobol indices in Figure \ref{fig:Sobol_Indices}b illustrates that $1/St$, $St$, $COT$, and $d_{eff}$ appear more sensitive to $p$ than $\mathrm{Re}_{out}$, while $\theta$ is more sensitive to $\mathrm{Re}_{out}$ than $p$. Therefore when discussing a model's sensitivity to a particular parameter, it can only be with respect to a specific performance metric. Moreover, the Sobol index rankings after $f$ among first-order indices are different for every output metric except $\theta$, see Figure \ref{fig:Sobol_Indices}a. Thus, as the first-order and total-order indices are not equivalent, this suggests higher-order interactions between parameters are important in the model output. With enough work, some local sensitivity estimates might be able to arrive at the same conclusion that a model is most sensitive to a particular parameter \cite{Link:2018}. However, if the goal of the sensitivity analysis is for model reduction, i.e., eliminating parameters or dynamics from a model, it might incorrectly suggest some parameters are not important, as it is difficult to predict the importance of parameter interactions using local sensitivity methods \cite{Saltelli:2019}.

The morphology and kinematics of the swimmer studied here resemble that of a \textit{C. elegan} \cite{Jung:2008}. This model has previously predicted swimming speeds ($1/St\sim0.4-0.7$) in agreement at lower biologically relevant Re ($\mathrm{Re}_{out}\sim0.5$) and frequencies ($f\sim 1.8-2.2\ \mathrm{Hz}$) \cite{Battista:ICB2020} with the organism itself ($1/St\sim0.53-0.75$) \cite{Jung:2010}. However, in this work the fluid scale investigated was approximately two or more orders of magnitude greater than that of a \textit{C. elegan} \cite{Jung:2010}. At higher Re, the emergent peak-to-peak stroke amplitude no longer appeared conserved, like that of \textit{C. elegans} when being placed in higher viscosity fluid environments \cite{Korta:2007}. The choice of Re in this work was deliberate to investigate this simple anguilliform mode further, over a range of Re that boasted the highest forward swimming speeds as observed in \cite{Battista:ICB2020}. This subset of input Reynolds numbers within the intermediate Re regime is a particularly interesting fluid scale to explore due to the importance and intricate balance of both viscous and inertial forces \cite{Jordan:1992,Tyson:2008}. Many of the seminal anguilliform studies have assumed either low Re \cite{Gray:1955,Gray:1964} or high Re \cite{Lighthill:1969} settings, and much of the work since has also focused on these two regimes, although a few studies have focused their efforts in this Re regime \cite{Tyson:2008,Gazzola:2012}. 

The C-start escape mechanism in larval fish falls within the intermediate Re regime ($Re\sim100s$) \cite{Muller:2008,Gazzola:2012}. However, it is not energetically efficient for the swimmer to continue swimming use this mode, rather they only use it to evade a predator before continuing with a less energy intensive anguilliform mode \cite{DuClos:2019}. During larval stages, some insects may use an anguilliform mode at $Re\sim100s$, such as Ceratopogonid larva, which has been observed swimming at an estimated $Re\sim160$ at $2.17$ bodylength/s \cite{Taylor:1952}. Other anguilliform swimmers, such as eels or lampreys, swim at higher Re in the 1000s or 10,000s \cite{Tytell:2004a,Tytell:2004,Tytell:2010}. However, their anguilliform modes are not equivalent to the simple swimmer studied here. A CFD model of lamprey swimming at $Re\sim1000s$ and $f=1\ \mathrm{Hz}$ resulted in swimming speeds between $0.25-0.5$ bodylengths/s \cite{Tytell:2010}. The model explored here produced swimmers within the range of $0.7-2.3$ bodylengths/s for swimmers operating at $\mathrm{Re}_{out}\sim100s$ (Figures \ref{fig:DIM_Colormap_Speed_COT} and \ref{fig:DIM_COT_vs_Speed} in the Supplemental Materials). Anguilliform swimmers such as American eels (\textit{Anguilla rostrata}), larval sea lampreys (\textit{Petromyzon marinus},~5-7 years old), and medicinal leeches (\textit{Hirudo medicinalis}) have been observed to swim at steady state speeds of $0.5-2$ bodylengths/s \cite{Tytell:2004,Tytell:2004c}, $1.6-1.75$ bodylengths/s \cite{DuClos:2019}, and $1.8-2$ bodylengths/s \cite{Jordan:1998}, respectively. As these aforementioned organisms all swim at high Re, it is important to note that hydrodynamics may substantially differ as Re changes from order 100s to 1000s (or beyond), as illustrated by differences in the flow fields of lamprey models \cite{Borazjani:2009,Tytell:2010b}.

Recently, much emphasis has been placed on designing optimal undulatory swimmers \cite{Kern:2006,Gazzola:2012,Tokic:2012,Eloy:2013,VanRees:2013,VanRees:2015,Gazzola:2015,Tokic:2019}. In general these studies attemptded to optimize either shape, kinematics, or both in regards to swimming speed and either efficiency or cost of transport. These ideas could help inform the design of faster, more efficient, more maneuverable underwater vehicles \cite{Low:2006,Niu:2014,Feng:2020}. Optimal morphologies and kinematics were different depending on whether maximal speed, maximal efficiency, or lowest cost of transport were desired \cite{Kern:2006,VanRees:2013,VanRees:2015,Tokic:2012,Tokic:2019}. Optimization studies by Toki\'c and Yue \cite{Tokic:2012,Tokic:2019} suggest that swimming speed and cost of transport are the main drivers of evolution, rather than locomotion efficiency. However, the studies above assumed high Re settings; only the studies by Gazzola et al. 2012 \cite{Gazzola:2012}, Van Rees et al. 2013 \cite{VanRees:2013}, and Van Rees et al. 2015 \cite{VanRees:2015} explored designs within the intermediate Re space - in particular, one specific Re, $\mathrm{Re}=550$. However, none of these studies varied Re across an intermediate regime nor did they vary undulation frequency. The global sensitivity results indicate that varying frequency across the intermediate Reynolds number regime of $50<\mathrm{Re}_{out}<1220$ will most significantly affect swimming performance, including both swimming speeds and cost of transports. On the other hand, this study did not alter the swimmer's morphology, i.e., its length, curvature, or thickness, nor properties of the stroke, i.e., stroke amplitude, resting time between successive strokes, or asymmetric stroke patterns. 

Furthermore, from the analysis in \cite{Battista:ICB2020}, it is evident that the parameter sensitivities quantified here may not remain true in other subspaces within the the entire $(\mathrm{Re}_{in},f,p)$ parameter space. It is important to note that changing the parameter space considered may lead to significant changes in the sensitivity results, e.g., rather than sampling $\mathrm{Re}_{in}\in[450,2200]$, selecting $\mathrm{Re}_{in}\in[10,100]$. For example, investigating the performance of the anguilliform swimming mode at different phases of development when Re may be smaller than $100$, might suggest a higher degree of sensitivity to Re rather than $f$, or other parameters entirely. Thus, to perform sensitivity analysis properly, insights from experimental data or other parameter explorations must first be carefully analyzed to make sure the analysis is performed over the correct parameter space.

Moreover, projections of the Sobol sampled $3D$ parameter space, $\{(\mathrm{Re}_{in_{j}},f_j,p_j)\}_{j=1}^{5000}$, were projected onto two-parameter subspaces, i.e., $(\mathrm{Re}_{in},f)$, $(\mathrm{Re}_{in},p)$, and $(f,p)$. Similar trends within the performance metrics could be recognized through these projections (Figures \ref{fig:Colormap_Speed_COT}, \ref{fig:Colormap_St_Deff}, and \ref{fig:DIM_Colormap_Speed_COT}) and the Pareto-like front identified (Figures \ref{fig:Results_COT_vs_Speed} and \ref{fig:DIM_COT_vs_Speed} as those observed in \cite{Battista:ICB2020}; however, there distinct differences between the two analyses.

First, in \cite{Battista:ICB2020}, the parameter combinations were selectively sampled in a rectangular fashion. Thus each subspace considered variations in two parameters at a time while holding the third parameter constant, rather than varying all three at once. The latter is what was done in the work here, via Sobol sequence sampling. The $3D$ data was then projected from a higher dimensional parameter subspace ($3D$) to a lower dimensional parameter subspace ($2D$) to investigate trends. Second, the projected colormaps here (Figures \ref{fig:Colormap_Speed_COT} and \ref{fig:Colormap_St_Deff}) divulge that the performance metrics appear most sensitive to changes in frequency, as most discernible qualitative patterns align with changes in $f$; however, each subspace also contained regions in which looked more like noise, which does not immediately suggest dominate parameter or parameter combinations. Third, while this work only projected the $3D$ performance data onto $2D$ subspaces, this sensitivity minded approach allowed us to also perform explicit global sensitivity analysis using Sobol sensitivity analysis. The broad parameter explorations in \cite{Battista:ICB2020} did not allow this; however, from the explicit $2D$ subspaces explored naive hypotheses could be formulated based on such explorations, but only in a strict qualitative sense. Fourth, note that the parameter space sampled in this paper was only realized from the explorations in \cite{Battista:ICB2020}, that is, a $3D$ subspace was first identified in which there was higher swimming performance in order to sub-sample for the sensitivity analysis performed here. Hence, in order to perform this analysis, such a region must be identified \textit{a priori}. Therefore parameter explorations are still necessary before performing sensitivity analysis if one is interested in the sensitivity to parameters within regions of desired performance, unless experimental data is readily available. If one changes the input parameter space sampled for sensitivity, the sensitivity indices, like those in Figure \ref{fig:Sobol_Indices} or \ref{fig:Sobol_Indices_2}, could significantly change. Thus, while Sobol sensitivity analysis gives global sensitivities of model output metrics to model parameters, the sensitivity results are highly dependent on the sampled parameter space \cite{Gan:2014,Zhang:2015}. 

Both the study presented here and \cite{Battista:ICB2020} only considered a $3D$ parameter space to analyze swimming performance over; however, the model's parameter space contains many more parameters that were not analyzed, such as those pertaining to the  swimmer's morphology (curvature and overall shape), asymmetric stroke patterns, or amount of rest between successive strokes. It took nearly 5000 simulations for the Sobol sensitivity indices to converge when studying a 3-dimensional space. Therefore studying a higher dimensional parameter space ($>3$) using this formulation will require exponentially more simulations \cite{Nossent:2011}. Contemporary methods involving polynomial chaos have become popular to reduce the computational burden, i.e., reduce the amount of simulations required for accurate indices \cite{Xiu:2003,Xiu:2003b,Sudret:2008}. However, to that extent, reducing the number of simulations bears the burden of not being able to resolve (potential) projected parameter subspaces that offer qualitative patterns and trends in the data as those here (Figures \ref{fig:Colormap_Speed_COT}, \ref{fig:Colormap_St_Deff}, \ref{fig:Colormap_RealRe_Freq}, \ref{fig:Colormap_RealRe_P} and \ref{fig:DIM_Colormap_Speed_COT}). 


Finally, when using mathematical modeling for biological inquiry, sensitivity analyses can help further knowledge of a biological system. It provides insight into the importance of parameters for specific output metrics. For example, if one wished to decipher which parameter (phenotype) was most important to the success of an active predation strategy, they would have to restrict the parameter space for a given model to a biologically relevant one for that particular organism. However, sensitivity insights alone aren't sufficient to fully understand any possible limitations of a particular predation strategy. Other parameter explorations may complement the sensitivity results to possibly inform why that organism has evolved into a particular parameter space, e.g., maybe such a strategy would not work (or be nearly as successful) if it grew to 10x its size (thus increasing the Re by a factor of 10). Both parameter explorations as well as sensitivity analyses can be used in-conjunction with one another to provide greater understanding a biological system. Symbiosis \footnote{only until you run out of computing time (environmental factors)}.

\bibliographystyle{naturemag}
\bibliography{Swim}

\begin{thebibliography}{10}
\expandafter\ifx\csname url\endcsname\relax
  \def\url#1{\texttt{#1}}\fi
\expandafter\ifx\csname urlprefix\endcsname\relax\def\urlprefix{URL }\fi
\providecommand{\bibinfo}[2]{#2}
\providecommand{\eprint}[2][]{\url{#2}}

\bibitem{Battista:ICB2020}
\bibinfo{author}{Battista, N.~A.}
\newblock \bibinfo{title}{Swimming through parameter subspaces of a simple
  swimmer}.
\newblock \emph{\bibinfo{journal}{Int. Comp. Biol.}}  (\bibinfo{year}{2020}).
\newblock \bibinfo{note}{Accepted, in production}.

\bibitem{Wu:2013}
\bibinfo{author}{Wu, J.}, \bibinfo{author}{Dhingra, R.},
  \bibinfo{author}{Gambhir, M.} \& \bibinfo{author}{Remais, J.~V.}
\newblock \bibinfo{title}{Sensitivity analysis of infectious disease models:
  methods, advances and their application}.
\newblock \emph{\bibinfo{journal}{J. R. Soc. Interface}}
  \textbf{\bibinfo{volume}{10(86)}}, \bibinfo{pages}{1020121018}
  (\bibinfo{year}{2013}).

\bibitem{BattistaBMB:2019}
\bibinfo{author}{Battista, N.~A.}, \bibinfo{author}{Pearcy, L.~B.} \&
  \bibinfo{author}{Strickland, W.~C.}
\newblock \bibinfo{title}{Modeling the prescription opioid epidemic}.
\newblock \emph{\bibinfo{journal}{Bull. Math. Biol.}}
  \textbf{\bibinfo{volume}{81(7)}}, \bibinfo{pages}{2258--2289}
  (\bibinfo{year}{2019}).

\bibitem{Lodhi:2011}
\bibinfo{author}{Lodhi, H.} \& \bibinfo{author}{Gilbert, D.}
\newblock \bibinfo{title}{Bootstrapping parameter estimation in dynamic
  systems}.
\newblock In \bibinfo{editor}{Elomaa, T.}, \bibinfo{editor}{J, H.} \&
  \bibinfo{editor}{Mannila, H.} (eds.) \emph{\bibinfo{booktitle}{Discovery
  Science}}, chap.~\bibinfo{chapter}{17}, \bibinfo{pages}{194--208}
  (\bibinfo{publisher}{Springer}, \bibinfo{address}{Berlin, Heidelberg,
  Germany}, \bibinfo{year}{2011}).

\bibitem{Barish:2017}
\bibinfo{author}{Barish, S.}, \bibinfo{author}{Ochs, M.~F.},
  \bibinfo{author}{Sontag, E.~D.} \& \bibinfo{author}{Gevertz, J.~L.}
\newblock \bibinfo{title}{Evaluating optimal therapy robustness by virtual
  expansion of a sample population, with a case study in cancer immunotherapy}.
\newblock \emph{\bibinfo{journal}{Proc. Natl. Acad. Sci.}}
  \textbf{\bibinfo{volume}{114}}, \bibinfo{pages}{E6277--E6286}
  (\bibinfo{year}{2017}).

\bibitem{Eriksson:2019}
\bibinfo{author}{Eriksson, O.} \emph{et~al.}
\newblock \bibinfo{title}{Uncertainty quantification, propagation and
  characterization by bayesian analysis combined with global sensitivity
  analysis applied to dynamical intracellular pathway models}.
\newblock \emph{\bibinfo{journal}{Bioinformatics}}
  \textbf{\bibinfo{volume}{35(2)}}, \bibinfo{pages}{284--292}
  (\bibinfo{year}{2019}).

\bibitem{Waldrop:2018}
\bibinfo{author}{Waldrop, L.~D.}, \bibinfo{author}{He, Y.} \&
  \bibinfo{author}{Khatri, S.}
\newblock \bibinfo{title}{What can computational modeling tell us about the
  diversity of odor-capture structures in the pancrustacea?}
\newblock \emph{\bibinfo{journal}{J. Chem. Ecol.}}
  \textbf{\bibinfo{volume}{44}}, \bibinfo{pages}{1084--1100}
  (\bibinfo{year}{2018}).

\bibitem{Waldrop:2020}
\bibinfo{author}{Waldrop, L.~D.}, \bibinfo{author}{He, Y.},
  \bibinfo{author}{Battista, N.~A.}, \bibinfo{author}{{Neary Peterman}, T.} \&
  \bibinfo{author}{Miller, L.~A.}
\newblock \bibinfo{title}{Uncertainty quantification reveals the physical
  constraints on pumping by peristaltic hearts}.
\newblock \emph{\bibinfo{journal}{Submitted}}  (\bibinfo{year}{2020}).

\bibitem{Saltelli:2002}
\bibinfo{author}{Saltelli, A.}
\newblock \bibinfo{title}{Making best use of model evaluations to compute
  sensitivity indices}.
\newblock \emph{\bibinfo{journal}{Comp. Phys. Comm.}}
  \textbf{\bibinfo{volume}{145}}, \bibinfo{pages}{280--297}
  (\bibinfo{year}{2002}).

\bibitem{Anderson:2015}
\bibinfo{author}{Anderson, P. S.~L.} \& \bibinfo{author}{Patek, S.~N.}
\newblock \bibinfo{title}{Mechanical sensitivity reveals evolutionary dynamics
  of mechanical systems}.
\newblock \emph{\bibinfo{journal}{Proceedings of the Royal Society B:
  Biological Sciences}} \textbf{\bibinfo{volume}{282}},
  \bibinfo{pages}{20143088} (\bibinfo{year}{2015}).

\bibitem{Munoz:2017}
\bibinfo{author}{Mu{\~n}oz, M.~M.}, \bibinfo{author}{Anderson, P. S.~L.} \&
  \bibinfo{author}{Patek, S.~N.}
\newblock \bibinfo{title}{Mechanical sensitivity and the dynamics of
  evolutionary rate shifts in biomechanical systems}.
\newblock \emph{\bibinfo{journal}{Proceedings of the Royal Society B:
  Biological Sciences}} \textbf{\bibinfo{volume}{284}},
  \bibinfo{pages}{20162325} (\bibinfo{year}{2017}).

\bibitem{Wainwright:2005}
\bibinfo{author}{Wainwright, P.~C.}, \bibinfo{author}{Alfaro, M.~E.},
  \bibinfo{author}{Bolnick, D.~I.} \& \bibinfo{author}{Hulsey, C.~D.}
\newblock \bibinfo{title}{{Many-to-One Mapping of Form to Function: A General
  Principle in Organismal Design?}}
\newblock \emph{\bibinfo{journal}{Integrative and Comparative Biology}}
  \textbf{\bibinfo{volume}{45}}, \bibinfo{pages}{256--262}
  (\bibinfo{year}{2005}).

\bibitem{Wainwright:2007}
\bibinfo{author}{Wainwright, P.~C.}
\newblock \bibinfo{title}{Functional versus morphological diversity in
  macroevolution}.
\newblock \emph{\bibinfo{journal}{Annual Review of Ecology, Evolution, and
  Systematics}} \textbf{\bibinfo{volume}{38}}, \bibinfo{pages}{381--401}
  (\bibinfo{year}{2007}).

\bibitem{Arnold:2003}
\bibinfo{author}{Arnold, S.~J.}
\newblock \bibinfo{title}{{Performance Surfaces and Adaptive Landscapes1}}.
\newblock \emph{\bibinfo{journal}{Integrative and Comparative Biology}}
  \textbf{\bibinfo{volume}{43}}, \bibinfo{pages}{367--375}
  (\bibinfo{year}{2003}).

\bibitem{Munoz:2018}
\bibinfo{author}{Mu{\~n}oz, M.~M.}, \bibinfo{author}{Hu, Y.},
  \bibinfo{author}{Anderson, P. S.~L.} \& \bibinfo{author}{Patek, S.}
\newblock \bibinfo{title}{Strong biomechanical relationships bias the tempo and
  mode of morphological evolution}.
\newblock \emph{\bibinfo{journal}{eLife}} \textbf{\bibinfo{volume}{7}},
  \bibinfo{pages}{e37621} (\bibinfo{year}{2018}).

\bibitem{Munoz:2019}
\bibinfo{author}{Mu{\~n}oz, M.~M.}
\newblock \bibinfo{title}{{The Evolutionary Dynamics of Mechanically Complex
  Systems}}.
\newblock \emph{\bibinfo{journal}{Integrative and Comparative Biology}}
  \textbf{\bibinfo{volume}{59}}, \bibinfo{pages}{705--715}
  (\bibinfo{year}{2019}).

\bibitem{Gray:1964}
\bibinfo{author}{Gray, J.} \& \bibinfo{author}{Lissmann, H.~W.}
\newblock \bibinfo{title}{The locomotion of nematodes}.
\newblock \emph{\bibinfo{journal}{J. Exp. Biol.}}
  \textbf{\bibinfo{volume}{41}}, \bibinfo{pages}{135--154}
  (\bibinfo{year}{1964}).

\bibitem{Ghosh:2008}
\bibinfo{author}{Ghosh, R.} \& \bibinfo{author}{Emmons, S.~W.}
\newblock \bibinfo{title}{Episodic swimming behavior in the nematode c.
  elegans}.
\newblock \emph{\bibinfo{journal}{J. Exp. Biol.}}
  \textbf{\bibinfo{volume}{211}}, \bibinfo{pages}{3703--3711}
  (\bibinfo{year}{2008}).

\bibitem{Majmudar:2012}
\bibinfo{author}{Majmudar, T.}, \bibinfo{author}{Keaveny, E.~E.},
  \bibinfo{author}{Zhang, J.} \& \bibinfo{author}{Shelley, M.~J.}
\newblock \bibinfo{title}{Experiments and theory of undulatory locomotion in a
  simple structured medium}.
\newblock \emph{\bibinfo{journal}{Journal of The Royal Society Interface}}
  \textbf{\bibinfo{volume}{9}}, \bibinfo{pages}{1809--1823}
  (\bibinfo{year}{2012}).

\bibitem{Luersen:2014}
\bibinfo{author}{L{\"u}ersen, K.}, \bibinfo{author}{Faust, U.},
  \bibinfo{author}{Gottschling, D.-C.} \& \bibinfo{author}{D{\"o}ring, F.}
\newblock \bibinfo{title}{Gait-specific adaptation of locomotor activity in
  response to dietary restriction in caenorhabditis elegans}.
\newblock \emph{\bibinfo{journal}{J. Exp. Biol.}}
  \textbf{\bibinfo{volume}{217}}, \bibinfo{pages}{2480--2488}
  (\bibinfo{year}{2014}).

\bibitem{Berman:2013}
\bibinfo{author}{Berman, R.~S.}, \bibinfo{author}{Kenneth, O.},
  \bibinfo{author}{Sznitman, J.} \& \bibinfo{author}{Leshansky, A.~M.}
\newblock \bibinfo{title}{Undulatory locomotion of finite filaments: lessons
  from \textit{Caenorhabditis elegans}}.
\newblock \emph{\bibinfo{journal}{New Journal of Physics}}
  \textbf{\bibinfo{volume}{15}}, \bibinfo{pages}{075022}
  (\bibinfo{year}{2013}).

\bibitem{Gutierrez:2014}
\bibinfo{author}{Gutierrez, J.}, \bibinfo{author}{Sorenson, M.} \&
  \bibinfo{author}{Strawbridge, E.}
\newblock \bibinfo{title}{Modeling fluid flow induced by c. elegans swimming at
  low reynolds number.}
\newblock In \bibinfo{editor}{Dediu, A.~H.}, \bibinfo{editor}{M, M.~L.} \&
  \bibinfo{editor}{{Mart\'in-Vide}, C.} (eds.) \emph{\bibinfo{booktitle}{Theory
  and Practice of Natural Computing}}, chap.~\bibinfo{chapter}{7},
  \bibinfo{pages}{71--82} (\bibinfo{publisher}{Springer},
  \bibinfo{address}{Cham, Switzerland}, \bibinfo{year}{2014}).

\bibitem{Montenegro:2016}
\bibinfo{author}{Montenegro-Johnson, T.~D.}, \bibinfo{author}{Gagnon, D.~A.},
  \bibinfo{author}{Arratia, P.~E.} \& \bibinfo{author}{Lauga, E.}
\newblock \bibinfo{title}{Flow analysis of the low reynolds number swimmer c.
  elegans}.
\newblock \emph{\bibinfo{journal}{Phys. Rev. Fluids}}
  \textbf{\bibinfo{volume}{1}}, \bibinfo{pages}{053202} (\bibinfo{year}{2016}).

\bibitem{Taylor:1952}
\bibinfo{author}{Taylor, G.~I.}
\newblock \bibinfo{title}{Analysis of the swimming of long and narrow animals}.
\newblock \emph{\bibinfo{journal}{Proc. Roy. Soc. London. Ser. A}}
  \textbf{\bibinfo{volume}{214}}, \bibinfo{pages}{158--183}
  (\bibinfo{year}{1952}).

\bibitem{Jordan:1998}
\bibinfo{author}{Jordan, C.~E.}
\newblock \bibinfo{title}{Scale effects in the kinematics and dynamics of
  swimming leeches}.
\newblock \emph{\bibinfo{journal}{Canadian Journal of Zoology}}
  \textbf{\bibinfo{volume}{76}}, \bibinfo{pages}{1869--1877}
  (\bibinfo{year}{1998}).

\bibitem{Tytell:2004}
\bibinfo{author}{Tytell, E.~D.}
\newblock \bibinfo{title}{The hydrodynamics of eel swimming ii. effect of
  swimming speed}.
\newblock \emph{\bibinfo{journal}{J. Exp. Biol.}}
  \textbf{\bibinfo{volume}{207}}, \bibinfo{pages}{3265--3279}
  (\bibinfo{year}{2004}).

\bibitem{Hamlet:2015}
\bibinfo{author}{Hamlet, C.}, \bibinfo{author}{Fauci, L.~J.} \&
  \bibinfo{author}{Tytell, E.~D.}
\newblock \bibinfo{title}{The effect of intrinsic muscular nonlinearities on
  the energetics of locomotion in a computational model of an anguilliform
  swimmer}.
\newblock \emph{\bibinfo{journal}{J. Theor. Biol.}}
  \textbf{\bibinfo{volume}{385}}, \bibinfo{pages}{119--129}
  (\bibinfo{year}{2015}).

\bibitem{Korta:2007}
\bibinfo{author}{Korta, J.}, \bibinfo{author}{Clark, D.~A.},
  \bibinfo{author}{Gabel, C.~V.}, \bibinfo{author}{Mahadevan, L.} \&
  \bibinfo{author}{Samuel, A. D.~T.}
\newblock \bibinfo{title}{Mechanosensation and mechanical load modulate the
  locomotory gait of swimming \textit{C. elegans}}.
\newblock \emph{\bibinfo{journal}{J. Exp. Biol.}}
  \textbf{\bibinfo{volume}{210}}, \bibinfo{pages}{2383--2389}
  (\bibinfo{year}{2007}).

\bibitem{Sobol:2001}
\bibinfo{author}{Sobol, I.~M.}
\newblock \bibinfo{title}{Global sensitivity indices for nonlinear mathematical
  models and their {M}onte {C}arlo estimates}.
\newblock \emph{\bibinfo{journal}{Math. and Comput. in Simul.}}
  \textbf{\bibinfo{volume}{55}}, \bibinfo{pages}{271--280}
  (\bibinfo{year}{2001}).

\bibitem{Eloy:2013}
\bibinfo{author}{Eloy, C.}
\newblock \bibinfo{title}{On the best design for undulatory swimming}.
\newblock \emph{\bibinfo{journal}{Journal of Fluid Mechanics}}
  \textbf{\bibinfo{volume}{717}}, \bibinfo{pages}{48–89}
  (\bibinfo{year}{2013}).

\bibitem{Verma:2017}
\bibinfo{author}{Verma, S.}, \bibinfo{author}{Hadjidoukas, P.},
  \bibinfo{author}{Wirth, P.}, \bibinfo{author}{Rossinelli, D.} \&
  \bibinfo{author}{Koumoutsakos, P.}
\newblock \bibinfo{title}{Pareto optimal swimmers}.
\newblock In \emph{\bibinfo{booktitle}{Proceedings of the Platform for Advanced
  Scientific Computing Conference}}, PASC ’17
  (\bibinfo{publisher}{Association for Computing Machinery},
  \bibinfo{address}{New York, NY, USA}, \bibinfo{year}{2017}).

\bibitem{Smits:2019}
\bibinfo{author}{Smits, A.~J.}
\newblock \bibinfo{title}{Undulatory and oscillatory swimming}.
\newblock \emph{\bibinfo{journal}{Journal of Fluid Mechanics}}
  \textbf{\bibinfo{volume}{874}}, \bibinfo{pages}{P1} (\bibinfo{year}{2019}).

\bibitem{BattistaIB2d:2018}
\bibinfo{author}{Battista, N.~A.}, \bibinfo{author}{Strickland, W.~C.},
  \bibinfo{author}{Barrett, A.} \& \bibinfo{author}{Miller, L.~A.}
\newblock \bibinfo{title}{{IB2d Reloaded: a more powerful Python and MATLAB
  implementation of the immersed boundary method}}.
\newblock \emph{\bibinfo{journal}{Math. Method. Appl. Sci}}
  \textbf{\bibinfo{volume}{41}}, \bibinfo{pages}{8455--8480}
  (\bibinfo{year}{2018}).

\bibitem{Battista:2020}
\bibinfo{author}{Battista, N.~A.}
\newblock \bibinfo{title}{Fluid-structure interaction for the classroom:
  Interpolation, hearts, and swimming!}
\newblock \emph{\bibinfo{journal}{(accepted, in production) SIAM Review}}
  (\bibinfo{year}{2020}).

\bibitem{Jung:2008}
\bibinfo{author}{Jung, S.}, \bibinfo{author}{Lee, S.} \&
  \bibinfo{author}{Samuel, A.}
\newblock \bibinfo{title}{Swimming c. elegans in a wet granular medium}.
\newblock \emph{\bibinfo{journal}{Chaos: An Interdisciplinary Journal of
  Nonlinear Science}} \textbf{\bibinfo{volume}{18}}, \bibinfo{pages}{041106}
  (\bibinfo{year}{2008}).

\bibitem{Rayner:1995}
\bibinfo{author}{Rayner, J.~M.}
\newblock \bibinfo{title}{Dynamics of the vortex wakes of flying and swimming
  vertebrates}.
\newblock \emph{\bibinfo{journal}{Symp. Soc. Exp, Biol.}}
  \textbf{\bibinfo{volume}{49}}, \bibinfo{pages}{131--155}
  (\bibinfo{year}{1995}).

\bibitem{Battista:2015}
\bibinfo{author}{Battista, N.~A.}, \bibinfo{author}{Baird, A.~J.} \&
  \bibinfo{author}{Miller, L.~A.}
\newblock \bibinfo{title}{A mathematical model and matlab code for
  muscle-fluid-structure simulations}.
\newblock \emph{\bibinfo{journal}{Integr. Comp. Biol.}}
  \textbf{\bibinfo{volume}{55(5)}}, \bibinfo{pages}{901--911}
  (\bibinfo{year}{2015}).

\bibitem{BattistaIB2d:2017}
\bibinfo{author}{Battista, N.~A.}, \bibinfo{author}{Strickland, W.~C.} \&
  \bibinfo{author}{Miller, L.~A.}
\newblock \bibinfo{title}{{IB2d: a Python and MATLAB implementation of the
  immersed boundary method}}.
\newblock \emph{\bibinfo{journal}{Bioinspir. Biomim.}}
  \textbf{\bibinfo{volume}{12(3)}}, \bibinfo{pages}{036003}
  (\bibinfo{year}{2017}).

\bibitem{Peskin:2002}
\bibinfo{author}{Peskin, C.~S.}
\newblock \bibinfo{title}{The immersed boundary method}.
\newblock \emph{\bibinfo{journal}{Acta Numerica}}
  \textbf{\bibinfo{volume}{11}}, \bibinfo{pages}{479--517}
  (\bibinfo{year}{2002}).

\bibitem{Kern:2006}
\bibinfo{author}{Kern, S.} \& \bibinfo{author}{Koumoutsakos, P.}
\newblock \bibinfo{title}{Simulations of optimized anguilliform swimming}.
\newblock \emph{\bibinfo{journal}{J. Exp. Biol.}}
  \textbf{\bibinfo{volume}{209}}, \bibinfo{pages}{4841--4857}
  (\bibinfo{year}{2006}).

\bibitem{VanRees:2015}
\bibinfo{author}{{van Rees}, W.~M.}, \bibinfo{author}{Gazzola, M.} \&
  \bibinfo{author}{Koumoutsakos, P.}
\newblock \bibinfo{title}{Optimal morphokinematics for undulatory swimmers at
  intermediate reynolds numbers}.
\newblock \emph{\bibinfo{journal}{Journal of Fluid Mechanics}}
  \textbf{\bibinfo{volume}{775}}, \bibinfo{pages}{178--188}
  (\bibinfo{year}{2015}).

\bibitem{Backholm:2015}
\bibinfo{author}{Backholm, M.}, \bibinfo{author}{Kasper, A. K.~S.},
  \bibinfo{author}{Schulman, R.~D.}, \bibinfo{author}{Ryu, W.~S.} \&
  \bibinfo{author}{Dalnoki-Veress, K.}
\newblock \bibinfo{title}{The effects of viscosity on the undulatory swimming
  dynamics of \textit{C. elegans}}.
\newblock \emph{\bibinfo{journal}{Physics of Fluids}}
  \textbf{\bibinfo{volume}{27}}, \bibinfo{pages}{091901}
  (\bibinfo{year}{2015}).

\bibitem{Jung:2010}
\bibinfo{author}{Jung, S.}
\newblock \bibinfo{title}{Caenorhabditis elegans swimming in a saturated
  particulate system}.
\newblock \emph{\bibinfo{journal}{Physics of Fluids}}
  \textbf{\bibinfo{volume}{22}}, \bibinfo{pages}{031903}
  (\bibinfo{year}{2010}).

\bibitem{Tytell:2004a}
\bibinfo{author}{Tytell, E.~D.} \& \bibinfo{author}{Lauder, G.~V.}
\newblock \bibinfo{title}{The hydrodynamics of eel swimming}.
\newblock \emph{\bibinfo{journal}{J. Exp. Biol.}}
  \textbf{\bibinfo{volume}{207}}, \bibinfo{pages}{1825--1841}
  (\bibinfo{year}{2004}).

\bibitem{Cui:2017}
\bibinfo{author}{Cui, Z.}, \bibinfo{author}{Gu, X.}, \bibinfo{author}{Li, K.}
  \& \bibinfo{author}{Jiang, H.}
\newblock \bibinfo{title}{Cfd studies of the effects of waveform on swimming
  performance of carangiform fish}.
\newblock \emph{\bibinfo{journal}{Appl. Sci.}} \textbf{\bibinfo{volume}{7}},
  \bibinfo{pages}{149} (\bibinfo{year}{2017}).

\bibitem{Dai:2018}
\bibinfo{author}{Dai, L.}, \bibinfo{author}{He, G.}, \bibinfo{author}{Zhang,
  X.} \& \bibinfo{author}{Zhang, X.}
\newblock \bibinfo{title}{Stable formations of self-propelled fish-like
  swimmers induced by hydrodynamic interactions}.
\newblock \emph{\bibinfo{journal}{J. Roy. Soc. Inter.}}
  \textbf{\bibinfo{volume}{15}}, \bibinfo{pages}{20180490}
  (\bibinfo{year}{2018}).

\bibitem{Taylor:2003}
\bibinfo{author}{Taylor, G.~K.}, \bibinfo{author}{Nudds, R.~L.} \&
  \bibinfo{author}{Thomas, A.~L.}
\newblock \bibinfo{title}{Flying and swimming animals cruise at a strouhal
  number tuned for high power efficiency}.
\newblock \emph{\bibinfo{journal}{Nature}} \textbf{\bibinfo{volume}{425}},
  \bibinfo{pages}{707--711} (\bibinfo{year}{2003}).

\bibitem{Eloy:2012}
\bibinfo{author}{Eloy, C.}
\newblock \bibinfo{title}{Optimal strouhal number for swimming animals}.
\newblock \emph{\bibinfo{journal}{Journal of Fluids and Structures}}
  \textbf{\bibinfo{volume}{30}}, \bibinfo{pages}{205 -- 218}
  (\bibinfo{year}{2012}).

\bibitem{Bale:2014}
\bibinfo{author}{Bale, R.}, \bibinfo{author}{Hao, M.}, \bibinfo{author}{Bhalla,
  A.} \& \bibinfo{author}{Patankar, N.~A.}
\newblock \bibinfo{title}{Energy efficiency and allometry of movement of
  swimming and flying animals}.
\newblock \emph{\bibinfo{journal}{Proc. Natl. Acad. Sci.}}
  \textbf{\bibinfo{volume}{111(21)}}, \bibinfo{pages}{7517–7521}
  (\bibinfo{year}{2014}).

\bibitem{Saltelli:2010}
\bibinfo{author}{Saltelli, A.} \emph{et~al.}
\newblock \bibinfo{title}{Variance based sensitivity analysis of model output.
  design and estimator for the total sensitivity index}.
\newblock \emph{\bibinfo{journal}{Comp. Phys. Comm.}}
  \textbf{\bibinfo{volume}{18}}, \bibinfo{pages}{259--270}
  (\bibinfo{year}{2010}).

\bibitem{Link:2018}
\bibinfo{author}{Link, K.~G.} \emph{et~al.}
\newblock \bibinfo{title}{A local and global sensitivity analysis of a
  mathematical model of coagulation and platelet deposition under flow}.
\newblock \emph{\bibinfo{journal}{PLOS ONE}} \textbf{\bibinfo{volume}{13(7)}},
  \bibinfo{pages}{e0200917} (\bibinfo{year}{2018}).

\bibitem{Saltelli:2019}
\bibinfo{author}{Saltelli, A.} \emph{et~al.}
\newblock \bibinfo{title}{Why so many published sensitivity analyses are false:
  A systematic review of sensitivity analysis practices}.
\newblock \emph{\bibinfo{journal}{Environ. Model. \& Software}}
  \textbf{\bibinfo{volume}{114}}, \bibinfo{pages}{29 -- 39}
  (\bibinfo{year}{2019}).

\bibitem{TCNJ:ELSA}
\bibinfo{author}{{The College of New Jersey}}.
\newblock \bibinfo{title}{Electronic laboratory for science \& analysis (elsa)}
  (\bibinfo{year}{2020}).
\newblock \urlprefix\url{https://docs.hpc.tcnj.edu/}.
\newblock \bibinfo{note}{Accessed Online; accessed 24 January 2020}.

\bibitem{Gillis:1996}
\bibinfo{author}{Gillis, G.~B.}
\newblock \bibinfo{title}{Undulatory locomotion in elongate aquatic
  vertebrates: Anguilliform swimming since sir james gray}.
\newblock \emph{\bibinfo{journal}{American Zoologist}}
  \textbf{\bibinfo{volume}{36}}, \bibinfo{pages}{656--665}
  (\bibinfo{year}{1996}).

\bibitem{Padmanabhan:2012}
\bibinfo{author}{Padmanabhan, V.} \emph{et~al.}
\newblock \bibinfo{title}{Locomotion of c. elegans: A piecewise-harmonic
  curvature representation of nematode behavior}.
\newblock \emph{\bibinfo{journal}{PLOS ONE}} \textbf{\bibinfo{volume}{7}},
  \bibinfo{pages}{1--11} (\bibinfo{year}{2012}).
\newblock \urlprefix\url{https://doi.org/10.1371/journal.pone.0040121}.

\bibitem{VanRees:2013}
\bibinfo{author}{{van Rees}, W.~M.}, \bibinfo{author}{Gazzola, M.} \&
  \bibinfo{author}{Koumoutsakos, P.}
\newblock \bibinfo{title}{Optimal shapes for anguilliform swimmers at
  intermediate reynolds numbers}.
\newblock \emph{\bibinfo{journal}{Journal of Fluid Mechanics}}
  \textbf{\bibinfo{volume}{722}}, \bibinfo{pages}{R3} (\bibinfo{year}{2013}).

\bibitem{Williams:2015}
\bibinfo{author}{Williams, T.~L.} \& \bibinfo{author}{McMillen, T.}
\newblock \bibinfo{title}{Strategies for swimming: explorations of the
  behaviour of a neuro-musculo-mechanical model of the lamprey}.
\newblock \emph{\bibinfo{journal}{Biology Open}} \textbf{\bibinfo{volume}{4}},
  \bibinfo{pages}{253--258} (\bibinfo{year}{2015}).

\bibitem{Schuech:2019}
\bibinfo{author}{Schuech, R.}, \bibinfo{author}{Hoehfurtner, T.},
  \bibinfo{author}{Smith, D.~J.} \& \bibinfo{author}{Humphries, S.}
\newblock \bibinfo{title}{Motile curved bacteria are pareto-optimal}.
\newblock \emph{\bibinfo{journal}{PNAS}} \textbf{\bibinfo{volume}{116(29)}},
  \bibinfo{pages}{14440--14447} (\bibinfo{year}{2019}).

\bibitem{Tokic:2019}
\bibinfo{author}{Toki\'c, G.} \& \bibinfo{author}{Yue, D.}
\newblock \bibinfo{title}{Energetics of optimal undulatory swimming organisms}.
\newblock \emph{\bibinfo{journal}{PLOS Computational Biology}}
  \textbf{\bibinfo{volume}{15}}, \bibinfo{pages}{1--25} (\bibinfo{year}{2019}).

\bibitem{Alben:2012c}
\bibinfo{author}{Alben, S.}, \bibinfo{author}{Witt, C.},
  \bibinfo{author}{Baker, T.~V.}, \bibinfo{author}{Anderson, E.} \&
  \bibinfo{author}{Lauder, G.~V.}
\newblock \bibinfo{title}{Dynamics of freely swimming flexible foils}.
\newblock \emph{\bibinfo{journal}{Physics of Fluids}}
  \textbf{\bibinfo{volume}{24}}, \bibinfo{pages}{051901}
  (\bibinfo{year}{2012}).

\bibitem{Shelton:2014}
\bibinfo{author}{Shelton, R.~M.}, \bibinfo{author}{Thornycroft, P. J.~M.} \&
  \bibinfo{author}{Lauder, G.~V.}
\newblock \bibinfo{title}{Undulatory locomotion of flexible foils as biomimetic
  models for understanding fish propulsion}.
\newblock \emph{\bibinfo{journal}{J. Exp. Biol.}}
  \textbf{\bibinfo{volume}{217}}, \bibinfo{pages}{2110--2120}
  (\bibinfo{year}{2014}).

\bibitem{Bainbridge:1958}
\bibinfo{author}{Bainbridge, R.}
\newblock \bibinfo{title}{The speed of swimming of fish as related to size and
  to the frequency and amplitude of the tail beat}.
\newblock \emph{\bibinfo{journal}{J. Exp. Biol.}}
  \textbf{\bibinfo{volume}{35}}, \bibinfo{pages}{109--133}
  (\bibinfo{year}{1958}).

\bibitem{Gray:1968}
\bibinfo{author}{Gray, J.}
\newblock \emph{\bibinfo{title}{Animal Locomotion (World Naturalist)}}
  (\bibinfo{publisher}{Weidenfeld and Nicolson}, \bibinfo{address}{London, UK},
  \bibinfo{year}{1968}).

\bibitem{Steinhausen:2005}
\bibinfo{author}{Steinhausen, M.~F.}, \bibinfo{author}{Steffensen, J.~F.} \&
  \bibinfo{author}{Andersen, N.~G.}
\newblock \bibinfo{title}{Tail beat frequency as a predictor of swimming speed
  and oxygen consumption of saithe (\textit{Pollachius virens}) and whiting
  (\textit{Merlangius merlangus}) during forced swimming}.
\newblock \emph{\bibinfo{journal}{Marine Biology}}
  \textbf{\bibinfo{volume}{148}}, \bibinfo{pages}{197--204}
  (\bibinfo{year}{2005}).

\bibitem{Hoover:2018}
\bibinfo{author}{Hoover, A.~P.}, \bibinfo{author}{Cortez, R.},
  \bibinfo{author}{Tytell, E.} \& \bibinfo{author}{Fauci, L.}
\newblock \bibinfo{title}{Swimming performance, resonance and shape evolution
  in heaving flexible panels}.
\newblock \emph{\bibinfo{journal}{J. Fluid. Mech.}}
  \textbf{\bibinfo{volume}{847}}, \bibinfo{pages}{386--416}
  (\bibinfo{year}{2018}).

\bibitem{Hamlet:2018}
\bibinfo{author}{Hamlet, C.~L.}, \bibinfo{author}{Hoffman, K.~A.},
  \bibinfo{author}{Tytell, E.~D.} \& \bibinfo{author}{Fauci, L.~J.}
\newblock \bibinfo{title}{The role of curvature feedback in the energetics and
  dynamics of lamprey swimming: A closed-loop model}.
\newblock \emph{\bibinfo{journal}{PLOS Computational Biology}}
  \textbf{\bibinfo{volume}{14}}, \bibinfo{pages}{1--29} (\bibinfo{year}{2018}).
\newblock \urlprefix\url{https://doi.org/10.1371/journal.pcbi.1006324}.

\bibitem{Jordan:1992}
\bibinfo{author}{Jordan, C.~E.}
\newblock \bibinfo{title}{A model of rapid-start swimming at intermediate
  reynolds number: undulatory locomotion in the chaetognath \textit{Sagitta
  elegans}}.
\newblock \emph{\bibinfo{journal}{J. Exp. Biol.}}
  \textbf{\bibinfo{volume}{163}}, \bibinfo{pages}{119--137}
  (\bibinfo{year}{1992}).

\bibitem{Tyson:2008}
\bibinfo{author}{Tyson, R.}, \bibinfo{author}{Jordan, C.~E.} \&
  \bibinfo{author}{Hebert, J.}
\newblock \bibinfo{title}{Modelling anguilliform swimming at intermediate
  reynolds number: A review and a novel extension of immersed boundary method
  applications}.
\newblock \emph{\bibinfo{journal}{Computer Methods in Applied Mechanics and
  Engineering}} \textbf{\bibinfo{volume}{197}}, \bibinfo{pages}{2105 -- 2118}
  (\bibinfo{year}{2008}).

\bibitem{Gray:1955}
\bibinfo{author}{Gray, J.} \& \bibinfo{author}{Hancock, G.~J.}
\newblock \bibinfo{title}{The propulsion of sea-urchin spermatozoa}.
\newblock \emph{\bibinfo{journal}{J. Exp. Biol.}}
  \textbf{\bibinfo{volume}{32}}, \bibinfo{pages}{802--814}
  (\bibinfo{year}{1955}).

\bibitem{Lighthill:1969}
\bibinfo{author}{Lighthill, M.~J.}
\newblock \bibinfo{title}{Hydromechanics of aquatic animal propulsion}.
\newblock \emph{\bibinfo{journal}{Annual Review of Fluid Mechanics}}
  \textbf{\bibinfo{volume}{1}}, \bibinfo{pages}{413--446}
  (\bibinfo{year}{1969}).

\bibitem{Gazzola:2012}
\bibinfo{author}{Gazzola, M.}, \bibinfo{author}{{van Rees}, W.~M.} \&
  \bibinfo{author}{Koumoutsakos, P.}
\newblock \bibinfo{title}{C-start: optimal start of larval fish}.
\newblock \emph{\bibinfo{journal}{Journal of Fluid Mechanics}}
  \textbf{\bibinfo{volume}{698}}, \bibinfo{pages}{5--18}
  (\bibinfo{year}{2012}).

\bibitem{Muller:2008}
\bibinfo{author}{M{\"u}ller, U.~K.}, \bibinfo{author}{van~den Boogaart, J.
  G.~M.} \& \bibinfo{author}{van Leeuwen, J.~L.}
\newblock \bibinfo{title}{Flow patterns of larval fish: undulatory swimming in
  the intermediate flow regime}.
\newblock \emph{\bibinfo{journal}{J. Exp. Biol.}}
  \textbf{\bibinfo{volume}{211}}, \bibinfo{pages}{196--205}
  (\bibinfo{year}{2008}).

\bibitem{DuClos:2019}
\bibinfo{author}{Du~Clos, K.~T.} \emph{et~al.}
\newblock \bibinfo{title}{Thrust generation during steady swimming and
  acceleration from rest in anguilliform swimmers}.
\newblock \emph{\bibinfo{journal}{J. Exp. Biol.}}
  \textbf{\bibinfo{volume}{222}} (\bibinfo{year}{2019}).

\bibitem{Tytell:2010}
\bibinfo{author}{Tytell, E.}, \bibinfo{author}{Hsu, C.},
  \bibinfo{author}{Williams, T.}, \bibinfo{author}{Cohen, A.} \&
  \bibinfo{author}{Fauci, L.}
\newblock \bibinfo{title}{Interactions between internal forces, body stiffness,
  and fluid environment in a neuromechanical model of lamprey swimming}.
\newblock \emph{\bibinfo{journal}{Proc. Natl. Acad. Sci.}}
  \textbf{\bibinfo{volume}{107}}, \bibinfo{pages}{19832--19837}
  (\bibinfo{year}{2010}).

\bibitem{Tytell:2004c}
\bibinfo{author}{Tytell, E.~D.}
\newblock \bibinfo{title}{Kinematics and hydrodynamics of linear acceleration
  in eels, \textit{Anguilla rostrata}}.
\newblock \emph{\bibinfo{journal}{Proceedings: Biological Sciences}}
  \textbf{\bibinfo{volume}{271}}, \bibinfo{pages}{2535--2540}
  (\bibinfo{year}{2004}).

\bibitem{Borazjani:2009}
\bibinfo{author}{Borazjani, I.} \& \bibinfo{author}{Sotiropoulos, F.}
\newblock \bibinfo{title}{Numerical investigation of the hydrodynamics of
  anguilliform swimming in the transitional and inertial flow regimes}.
\newblock \emph{\bibinfo{journal}{J. Exp. Biol.}}
  \textbf{\bibinfo{volume}{212}}, \bibinfo{pages}{576--592}
  (\bibinfo{year}{2009}).

\bibitem{Tytell:2010b}
\bibinfo{author}{Tytell, E.~D.} \emph{et~al.}
\newblock \bibinfo{title}{Disentangling the functional roles of morphology and
  motion in the swimming of fish}.
\newblock \emph{\bibinfo{journal}{Int. Comp. Biol.}}
  \textbf{\bibinfo{volume}{50(6)}}, \bibinfo{pages}{1140–1154}
  (\bibinfo{year}{2010}).

\bibitem{Tokic:2012}
\bibinfo{author}{Toki\'c, G.} \& \bibinfo{author}{Yue, D.}
\newblock \bibinfo{title}{Optimal shape and motion of undulatory swimming
  organisms}.
\newblock \emph{\bibinfo{journal}{Proceedings of the Royal Society B:
  Biological Sciences}} \textbf{\bibinfo{volume}{279}},
  \bibinfo{pages}{3065--3074} (\bibinfo{year}{2012}).

\bibitem{Gazzola:2015}
\bibinfo{author}{Gazzola, M.}, \bibinfo{author}{Argentina, M.} \&
  \bibinfo{author}{Mahadevan, L.}
\newblock \bibinfo{title}{Gait and speed selection in slender inertial
  swimmers}.
\newblock \emph{\bibinfo{journal}{Proceedings of the National Academy of
  Sciences}} \textbf{\bibinfo{volume}{112}}, \bibinfo{pages}{3874--3879}
  (\bibinfo{year}{2015}).

\bibitem{Low:2006}
\bibinfo{author}{{Low}, K.~H.}, \bibinfo{author}{{Prabu}, S.} \&
  \bibinfo{author}{{Pattathil}, A.~P.}
\newblock \bibinfo{title}{Initial prototype design and development of hybrid
  modular underwater vehicles}.
\newblock In \emph{\bibinfo{booktitle}{2006 IEEE International Conference on
  Robotics and Biomimetics}}, \bibinfo{pages}{311--316} (\bibinfo{year}{2006}).

\bibitem{Niu:2014}
\bibinfo{author}{Niu, X.} \& \bibinfo{author}{Xu, J.}
\newblock \bibinfo{title}{Modeling, control and locomotion planning of an
  anguilliform robotic fish}.
\newblock \emph{\bibinfo{journal}{Unmanned Systems}}
  \textbf{\bibinfo{volume}{02}}, \bibinfo{pages}{295--321}
  (\bibinfo{year}{2014}).

\bibitem{Feng:2020}
\bibinfo{author}{Feng, H.}, \bibinfo{author}{Sun, Y.}, \bibinfo{author}{Todd,
  P.~A.} \& \bibinfo{author}{Lee, H.~P.}
\newblock \bibinfo{title}{Body wave generation for anguilliform locomotion
  using a fiber-reinforced soft fluidic elastomer actuator array toward the
  development of the eel-inspired underwater soft robot}.
\newblock \emph{\bibinfo{journal}{Soft Robotics}} \textbf{\bibinfo{volume}{7}},
  \bibinfo{pages}{233--250} (\bibinfo{year}{2020}).

\bibitem{Gan:2014}
\bibinfo{author}{Gan, Y.} \emph{et~al.}
\newblock \bibinfo{title}{A comprehensive evaluation of various sensitivity
  analysis methods: A case study with a hydrological model}.
\newblock \emph{\bibinfo{journal}{Envir. Model. Soft.}}
  \textbf{\bibinfo{volume}{51}}, \bibinfo{pages}{269--285}
  (\bibinfo{year}{2014}).

\bibitem{Zhang:2015}
\bibinfo{author}{Zhang, X.~Y.}, \bibinfo{author}{Trame, M.},
  \bibinfo{author}{Lesko, L.} \& \bibinfo{author}{Schmidt, S.}
\newblock \bibinfo{title}{Sobol sensitivity analysis: A tool to guide the
  development and evaluation of systems pharmacology models}.
\newblock \emph{\bibinfo{journal}{CPT Pharmacometrics Syst. Pharmacol.}}
  \textbf{\bibinfo{volume}{4}}, \bibinfo{pages}{69--79} (\bibinfo{year}{2014}).

\bibitem{Nossent:2011}
\bibinfo{author}{Nossent, J.}, \bibinfo{author}{Elsen, P.} \&
  \bibinfo{author}{Bauwens, W.}
\newblock \bibinfo{title}{Sobol’ sensitivity analysis of a complex
  environmental model}.
\newblock \emph{\bibinfo{journal}{Envir. Model. Soft.}}
  \textbf{\bibinfo{volume}{26(12)}}, \bibinfo{pages}{1515--1525}
  (\bibinfo{year}{2011}).

\bibitem{Xiu:2003}
\bibinfo{author}{Xiu, D.} \& \bibinfo{author}{Karniadakis, G.~E.}
\newblock \bibinfo{title}{Modeling uncertainty in flow simulations via
  generalized polynomial chaos}.
\newblock \emph{\bibinfo{journal}{J. Comp. Phys.}}
  \textbf{\bibinfo{volume}{187}}, \bibinfo{pages}{137--167}
  (\bibinfo{year}{2003}).

\bibitem{Xiu:2003b}
\bibinfo{author}{Xiu, D.}, \bibinfo{author}{Lucor, D.}, \bibinfo{author}{Su,
  C.~H.} \& \bibinfo{author}{Karniadakis, G.~E.}
\newblock \bibinfo{title}{Performance evaluation of generalized polynomial
  chaos}.
\newblock In \bibinfo{editor}{Sloot, P.} \emph{et~al.} (eds.)
  \emph{\bibinfo{booktitle}{In International Conference on Computational
  Science}}, chap.~\bibinfo{chapter}{36}, \bibinfo{pages}{346--354}
  (\bibinfo{publisher}{Springer}, \bibinfo{address}{Berlin, Heidelberg,
  Germany}, \bibinfo{year}{2003}).

\bibitem{Sudret:2008}
\bibinfo{author}{Sudret, B.}
\newblock \bibinfo{title}{Global sensitivity analysis using polynomial chaos
  expansions}.
\newblock \emph{\bibinfo{journal}{Reliability Engineering \& System Safety}}
  \textbf{\bibinfo{volume}{93(7)}}, \bibinfo{pages}{964--979}
  (\bibinfo{year}{2008}).

\end{thebibliography}

%
%



\begin{notes}[Acknowledgements]
The author would like to thank Lindsay Waldrop and Jonathan Rader for the invitation to participate in the SICB Symposium \textit{Melding Modeling and Morphology: integrating approaches to understand the evolution of form and function symposium} at the 2020 annual meeting. The author would also like to thank Christina Battista, Karen Clark, Jana Gevertz, Laura Miller, Matthew Mizuhara, Emily Slesinger, Edward Voskanian, and Lindsay Waldrop for comments and discussion. Computational resources were provided by the NSF OAC \#1826915 and the NSF OAC \#1828163. Support for N.A.B. was provided by the TCNJ Support of Scholarly Activity Grant, the TCNJ Department of Mathematics and Statistics, and the TCNJ School of Science.
\end{notes}


%
%

\newpage
\setcounter{page}{1}

\section*{Supplemental Materials}
\beginsupplement

%
%

\subsection*{Computational Parameters and Geometry}
\label{app:params}

Table \ref{table:num_param} offers the computational parameters used in the study. They are identical to those in \cite{Battista:ICB2020} with the exception of the parameter ranges for $\mathrm{Re}_{in}$, $f$, and $p$.

\begin{table}[H]
\begin{center}
\begin{tabular}{| c | c | c | c |}
    \hline
    Parameter               & Variable    & Units        & Value \\ \hline
    Domain Size            & $[L_x,L_y]$  & m               &  $[6,16]$             \\ \hline
    Spatial Grid Size      & $dx=dy$      & m               &  $L_x/1024=L_y/384$            \\ \hline
    Lagrangian Grid Size    & $ds$        & m               &  $dx/2$               \\ \hline
    Time Step Size          & $dt$        & s               &  $2.5\times10^{-5}$   \\ \hline
    Total Simulation Time    & $T$        & \textit{stroke cycles} &  $6$               \\ \hline
    Fluid Density            & $\rho$     & $kg/m^3$        &  $1000$               \\ \hline
    Fluid Dynamic Viscosity & $\mu$      & $kg/(ms)$       &  [0.475,17750]       \\ \hline
    Swimmer Length           & $L$        & m               &  $1.5$    \\ \hline
    Swimmer Height           & $h$        & m               &  $0.5$        \\ \hline
    Stroke Frequency         & $f$        & $s^{-1}$       &   [1,3.0]  \\ \hline
    Reynolds number (input)        & $\mathrm{Re}_{in}$        & -              &  [450,2200] \\ \hline
    Kinematic Parameter     & $p$        & $-$            &  [0.05,0.45] \\ \hline
    Spring Stiffness   & $k_{spr}$ & $kg\cdot m/s^2$ &  $9.5625\times10^{9}$  \\ \hline
    Non-invariant Beam Stiffness   & $k_{beam} $ & $kg\cdot m/s^2$ &  $2.03634\times10^{12}$  \\ \hline
    \end{tabular}
    \caption{Numerical parameters used in the two-dimensional immersed boundary simulations of the idealized, simple anguilliform swimmer}
    \label{table:num_param}
    \end{center}
\end{table}

The undulations of the swimmer's body that produce forward swimming are governed by interpolating between two curvature states. The function that governs the interpolation is given by a cubic spline \cite{Battista:2020,Battista:ICB2020}. This interpolant satisfies the following criteria:
\begin{enumerate}
    \item Continuity of the interpolation polynomial, its velocity, and its acceleration between successive half-strokes and stroke cycles
    \item No instantaneous acceleration or deceleration
    \item Symmetry within each half-stroke cycle
\end{enumerate}

\begin{figure}[H]
    \centering
    \includegraphics[width=0.90\textwidth]{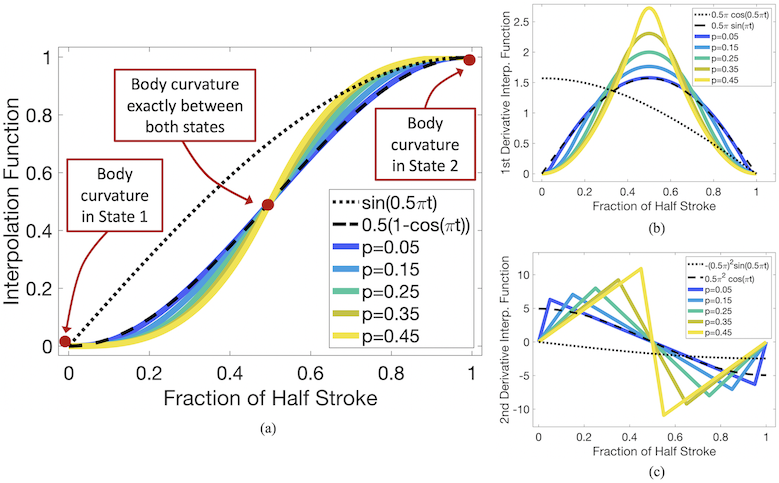}
    \caption{(a) Examples of how the interpolation function deviates away from a trivial sinusoidal interpolation between body curvature states as $p$ increases. The associated velocity and acceleration profiles are given in (b) and (c) respectively. Figure modified and courtesy of \cite{Battista:ICB2020}.}
    \label{fig:interpolation}
\end{figure}

The resulting interpolant can be seen in Figure \ref{fig:interpolation} for a variety of kinematic parameter ($p$) values. This figure was modified from \cite{Battista:ICB2020}. As $p$ increases, the interpolant deviates from trivial sinusoidal curvature interpolation functions. Some anguilliform studies investigating optimal morphokinematics use a mix of cubic splines and sinusoidal functions to prescribe the swimmer's body kinematics \cite{Kern:2006,VanRees:2015}; however, these constructions result in many more free parameters that govern the body's shape and kinematics, rather than only one $(p)$ in the simplified model explored here.

\begin{figure}[H]
    \centering
    \includegraphics[width=0.975\textwidth]{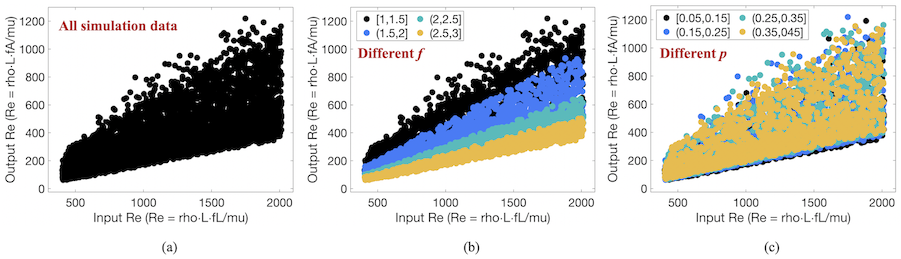}
    \caption{The input Reynolds number, $Re=\rho L(fL)/\mu$, plotted against the output Reynolds number, $Re=\rho L(fA)/\mu$, whose frequency based velocity scale ($fA$) is based off the undulation peak-to-peak amplitude, $A$, which is an output of the model.}
    \label{fig:input_Re_vs_output_Re}
\end{figure}

Since it was desired to have a fluid scaling parameter as an input to the model, an input Reynolds number was defined in Eq. \ref{eq:Re}, based on a characteristic length that was the swimmer's body ($L$) and a frequency based characteristic velocity that was the product of the undulation frequency and the swimmer's body length, $fL$ \cite{Cui:2017,Dai:2018}. Many anguilliform swimming studies use the characteristic velocity as $fA$, where $A$ is the peak-to-peak undulation amplitude. However, since $A$ is an output of the model, using a characteristic velocity scale of $fA$ was impossible for the input Re. Figure \ref{fig:input_Re_vs_output_Re} provides plots of the input Re vs. output Re, as organized by the input parameters $f$ and $p$ in Figures \ref{fig:input_Re_vs_output_Re}b and c, respectively. The output Re, $\mathrm{Re}_{out}$ depends on model output $A$, as a pattern emerges, where higher $f$ corresponds to lower $\mathrm{Re}_{out}$, given a particular $\mathrm{Re}_{in}$. This occurs as $f$ and $A$ are inversely related, i.e., higher $f$ results in a smaller $A$, see Figure \ref{fig:Colormap_St_Deff}).

%
%

\subsection*{Additional Data}
\label{app:additional_data}

\subsubsection*{Non-dimensional Data}

\begin{figure}[H]
    \centering
    \includegraphics[width=0.95\textwidth]{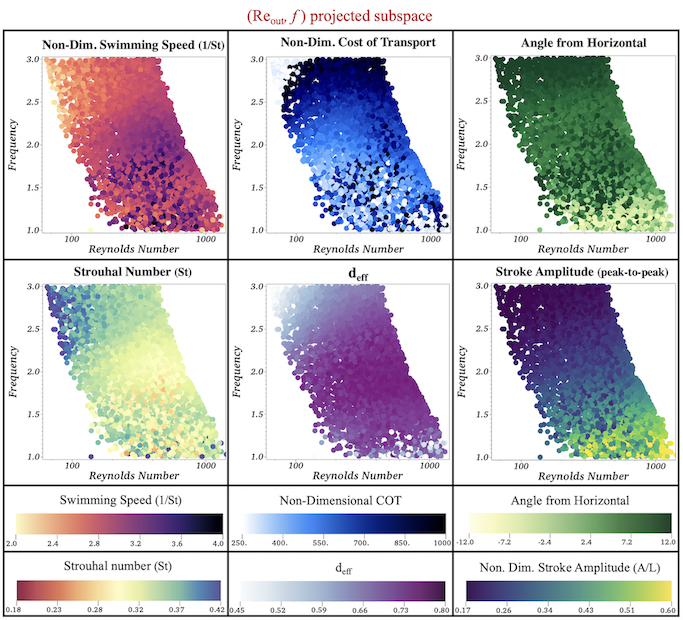}
    \caption{Colormaps corresponding to the $(\mathrm{Re}_{out},f)$ projected subspace, providing the non-dimensional forward swimming speeds ($1/St$), cost of transports ($COT$), and stroke amplitudes ($A/L$), as well as the Strouhal numbers ($St$), the distance effectiveness ratios ($d_{eff}$), and the angular trajectories off the horizontal ($\theta$) for all the data sampled from Sobol sequences. Note that the Reynolds numbers shown here are the $Re_{out}$, whose frequency based velocity scale ($fA$) is given in terms of the output stroke amplitude, $A$.}
    \label{fig:Colormap_RealRe_Freq}
\end{figure}

\begin{figure}[H]
    \centering
    \includegraphics[width=0.95\textwidth]{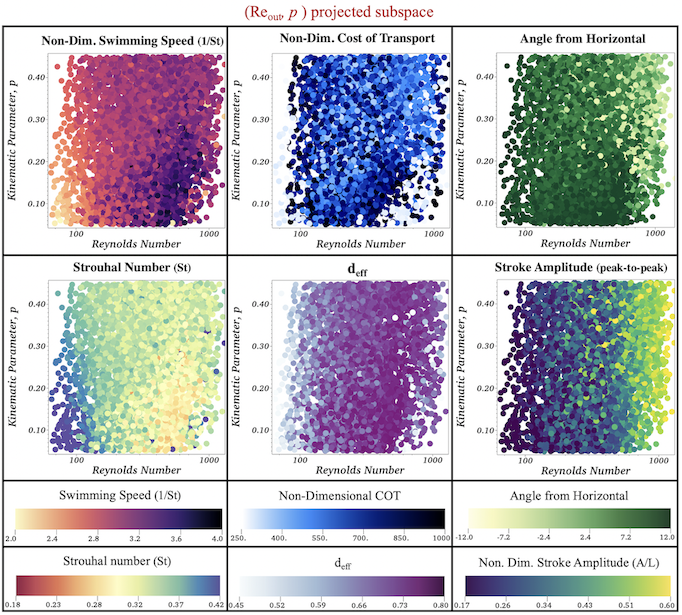}
    \caption{Colormaps corresponding to the $(\mathrm{Re}_{out},p)$ projected subspace, providing the non-dimensional forward swimming speeds ($1/St$), cost of transports ($COT$), and stroke amplitudes ($A/L$), as well as the Strouhal numbers ($St$), the distance effectiveness ratios ($d_{eff}$), and the angular trajectories off the horizontal ($\theta$) for all the data sampled from Sobol sequences. Note that the Reynolds numbers shown here are the $Re_{out}$, whose frequency based velocity scale ($fA$) is given in terms of the output stroke amplitude, $A$.}
    \label{fig:Colormap_RealRe_P}
\end{figure}

Figures \ref{fig:Colormap_RealRe_Freq} and \ref{fig:Colormap_RealRe_P} provide colormaps of the performance data in terms of the output Reynolds number, $Re_{out}=\rho\cdot L\cdot fA/\mu,$ based on the output peak-to-peak stroke amplitude, across the $(\mathrm{Re}_{in},f)$ and $(\mathrm{Re}_{in},p)$ projected subspaces, respectively. The resulting output Reynolds numbers spanned $[50,1220]$. Patterns emerge across both projected subspaces as either $\mathrm{Re}_{out}$ and either $f$ or $p$ vary in each subspace. Although, there are regions within the subspaces that for particular performance metrics look qualitatively like noise, e.g., the colormap for the $(\mathrm{Re}_{out},p)$ subspace for $COT$ when $p\gtrsim0.25$ for all $\mathrm{Re}_{out}$. Such noisy regions might suggest that the two varying subspace parameters are not the main parameters driving that performance metric in that region. The third free parameter (and possible interactions with the others) may have a more significant impact on that performance metric there.

Interestingly, both subspaces suggest that the largest regions of highest swimming speed are located within regions that may encompass both high and low values in $COT$, near $f\sim 2.0\ \mathrm{Hz}$. These regions also correspond to higher $d_{eff}$. Moreover, the Strouhal numbers almost everywhere within the subspaces fall within the optimal region of $0.2<St<0.4$ \cite{Taylor:2003}, except near the lower end of the $Re_{out}$ spectrum, where they begin to rise greater than 0.4 as $Re_{out}$ decreases.

Furthermore, the $(\mathrm{Re},f)$ projected parameter subspace data is qualitatively very similar when organized by either $\mathrm{Re}_{in}$ (Figures \ref{fig:Colormap_Speed_COT} and \ref{fig:Colormap_St_Deff}) or $\mathrm{Re}_{out}$ (Figure \ref{fig:Colormap_RealRe_Freq}). These consistencies suggest that the Sobol sensitivity analysis would remain consistent for either $\mathrm{Re}_{in}$ or $\mathrm{Re}_{out}$. However, there are clear differences in the case of $(\mathrm{Re},p)$ among Figures \ref{fig:Colormap_Speed_COT} and \ref{fig:Colormap_St_Deff}) when compared to Figure \ref{fig:Colormap_RealRe_P}. That is, the data appears noisy in the former, while patterns form in the latter. The patterns that emerge in Figure \ref{fig:Colormap_RealRe_P} indicate a strong dependence on $fA$, as it organized by $\mathrm{Re}_{out}$. However, this is actually to be expected since $\mathrm{Re}_{out}$ depends of the output $A$ and it is observed that $f$ and $A$ are inversely related (see the $A$ panel for the $(\mathrm{Re},f)$ and $(f,p)$ subspaces in Figure \ref{fig:Colormap_St_Deff}). So the patterns that emerge are a result that although the input frequency $f$ is not one of the axes, its presence is felt by the dependence of $A$ in $\mathrm{Re}_{out}$.

\subsubsection*{Dimensional Data}

\begin{figure}[H]
    \centering
    \includegraphics[width=0.90\textwidth]{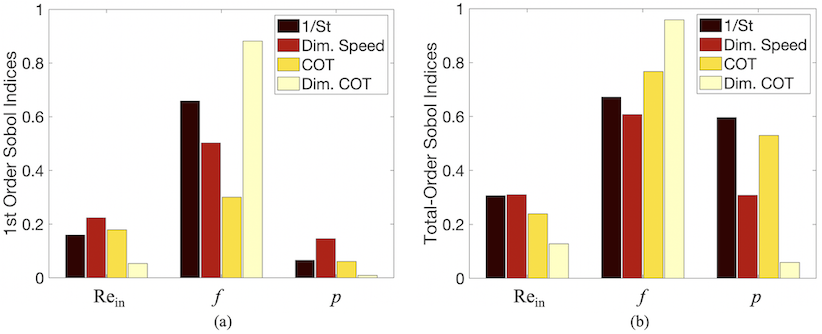}
    \caption{(a) First-order and (b) Total-order Sobol indices of the three varied parameters $\mathrm{Re}_{in}$, $f$, and $p$ to compare global parameter sensitivities of dimensional to non-dimensional quantities for swimming speed and cost of transport.}
    \label{fig:Sobol_Indices_2}
\end{figure}

A comparison of the global parameter sensitivities for swimming speed and cost of transport between their dimensional and non-dimensional form is provided in Figure \ref{fig:Sobol_Indices_2}. The dimensional output metrics for swimming speed (bodylength/s) and COT (N/kg) are still most sensitive to variations in the stroke (undulation) frequency within the $\mathrm{Re}_{in}\times f\times p = [450,2200]\times[1,3]\times[0.05,0.45]$ parameter subspace. However, both of the dimensional swimming speed and cost of transport are slightly more sensitive to $\mathrm{Re}_{in}$ than $p$.

\begin{figure}[H]
    \centering
    \includegraphics[width=0.95\textwidth]{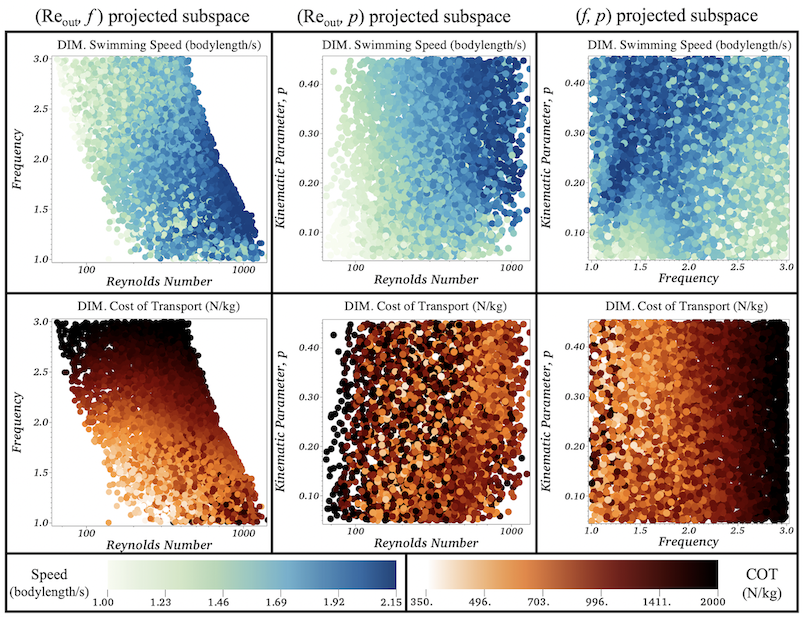}
    \caption{Colormaps corresponding to the \textit{dimensional} forward swimming speeds (bodylength/s) and cost of transports (N/kg) for all the data sampled from Sobol sequences when projected onto either the $(\mathrm{Re}_{out},f)$ and $(\mathrm{Re}_{out},p)$ subspaces.}
    \label{fig:DIM_Colormap_Speed_COT}
\end{figure}

Colormaps showing the \textit{dimensional} forward swimming speeds (bodylength/s) and cost of transports (N/kg) over each projected parameter subspace are provided in Figure \ref{fig:DIM_Colormap_Speed_COT}. Similar to Figures \ref{fig:Colormap_RealRe_Freq} and \ref{fig:Colormap_RealRe_P}, patterns emerge within the subspaces as different parameters are varied. Note that these Figures give the Reynolds number as $\mathrm{Re}_{out}$, unlike Figures \ref{fig:Colormap_Speed_COT} and \ref{fig:Colormap_St_Deff} in the main manuscript. Beyond similar trends in $f$, dimensional swimming speed appears to be highly correlated with higher $\mathrm{Re}_{out}$ and higher $p$. As Figure \ref{fig:interpolation} suggests, higher $p$ correspond to slower initial accelerations, but higher maximal velocities and accelerations of the changing body curvature throughout each half-stroke.

\begin{figure}[H]
    \centering
    \includegraphics[width=0.825\textwidth]{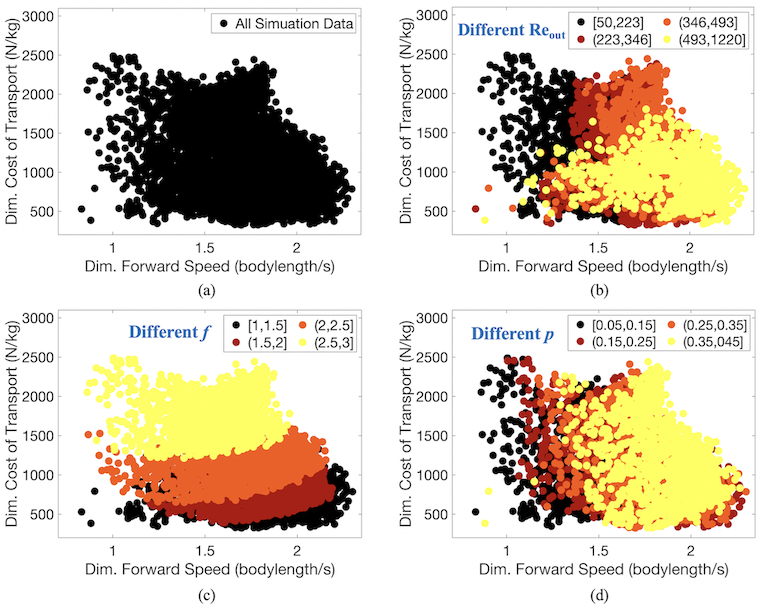}
    \caption{The \textit{dimensional} cost of transport (N/kg) and forward swimming speeds (bodylength/s) plotted against each other for (a) every simulation performed, as well as indicating clusters corresponding to differing ranges of the input parameters: (b) different $\mathrm{Re}_{out}$ (c) different $f$, and (d) different $p$.}
    \label{fig:DIM_COT_vs_Speed}
\end{figure}

Pareto-like optimal fronts were identified by plotting the \textit{dimensional} cost of transport (N/kg) against swimming speed (bodylength/s) for each simulation performed. The data is presented in Figure \ref{fig:DIM_COT_vs_Speed}. From Figure \ref{fig:DIM_COT_vs_Speed}d, given a $p$, depending on values of the other two parameters, one could construct a swimmer that falls almost anywhere within the performance space. However, Figure \ref{fig:DIM_COT_vs_Speed}c shows distinct clusters where different frequency ranges reside and Figure \ref{fig:DIM_COT_vs_Speed}b suggests higher $\mathrm{Re}_{out}$ result in faster swimming speeds (bodylength/s) and lower cost of transport (N/kg). This complemented the Sobol sensitivity results for the dimensional data analyzed; choosing a particular $f$ would most significantly determine where in the performance space a swimmer may reside. 

\begin{figure}[H]
    \centering
    \includegraphics[width=0.925\textwidth]{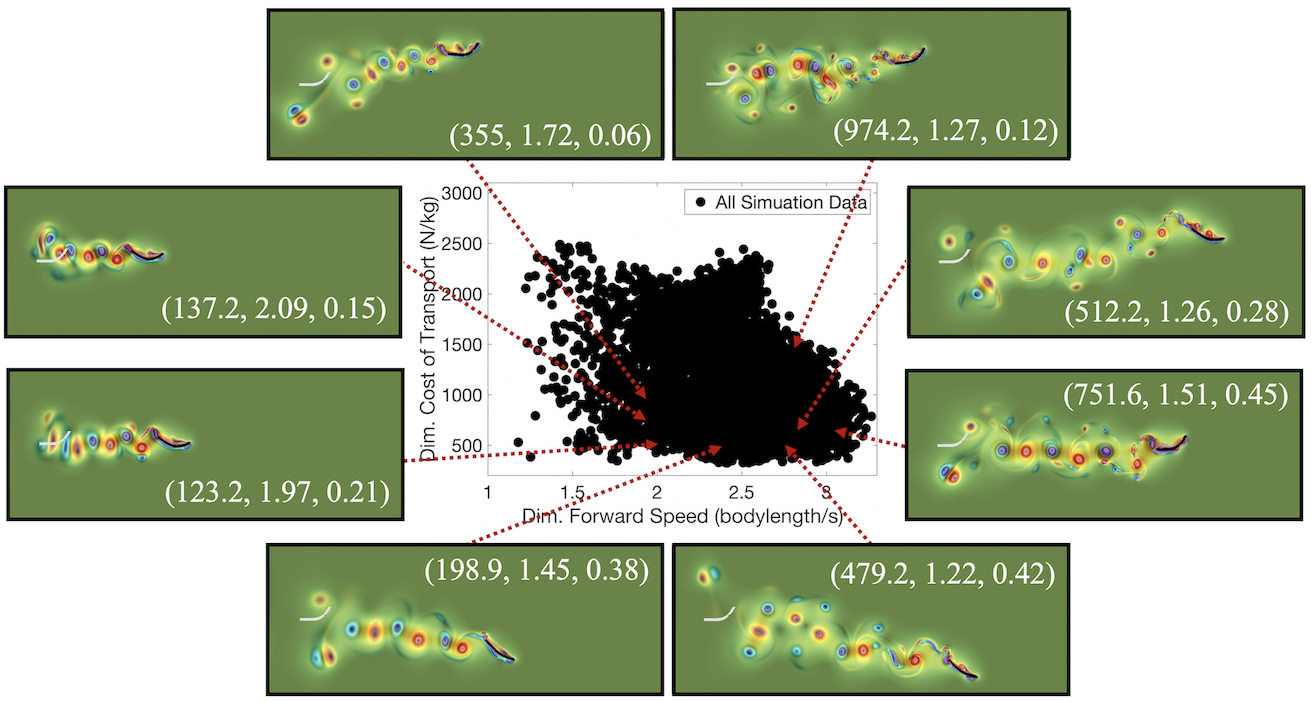}
    \caption{Numerous swimmer's vortex wake and position after its $5^{th}$ full stroke cycle as well as indicating where they fall onto the \textit{dimensional} cost of transport (N/kg) vs. swimming speed (bodylength/s) performance space. The colormap illustrates vorticity and the greyed out swimmer shows the starting position of the swimmer in each case. Different parameter combinations $(\mathrm{Re}_{out},f,p)$ lead to different swimming behavior, as indicated by its placement within the performance space and vortex wake left behind. }
    \label{fig:Pareto_Vorticity_Dim}
\end{figure}

Figure \ref{fig:Pareto_Vorticity_Dim} shows where different parameter combination swimmers lie within the performance space in Figure \ref{fig:Pareto_Vorticity_Dim}. This figure also illustrates those swimmer's position and vortex wakes after their $5^{th}$ stroke cycle. Again, the Reynolds number indicated here is the output Re, $\mathrm{Re}_{out}=\rho L(fA)/\mu,$ where $A$ is the peak-to-peak amplitude, computed as output from the model. These are the same swimmers illustrated in Figure \ref{fig:Pareto_Vorticity_NonDim}, but in the context of the dimensional performance space.

%
%

\subsection*{Details regarding the Immersed Boundary Method (IB)}
\label{app:IB}

Here the \textit{immersed boundary method} (IB) \cite{Peskin:2002} will be briefly introduced. The IB was fluid-structure interaction method that solved the equations that coupled the angulliform swimmer and the fluid to which it was immersed. The open-source IB software \texttt{IB2d} \cite{Battista:2015,BattistaIB2d:2017,BattistaIB2d:2018} was used for all the simulations presented in this work.

The full viscous, incompressible Navier-Stokes equations were used to model the fluid since $Re\in[250,750]$, i.e., 
\begin{equation} 
   \rho\Big[\frac{\partial\textbf{u}}{\partial t}({\textbf{x} },t) +\textbf{u}({\textbf {x}},t)\cdot\nabla \textbf{u}({\textbf{x}} ,t)\Big]=  -\nabla p({\textbf{x}},t) + \mu \Delta \textbf{u}({\textbf{x}},t) + \textbf{F}({\textbf{x}},t) \label{eq:NS1}
\end{equation}
\begin{equation}
      \nabla\cdot \textbf{u}({\bf x},t) = 0 \label{eq:NSDiv1}
\end{equation}
where $\textbf{u}({\textbf{x}},t)$ and $p({\textbf{x}},t)$are the fluid's velocity and pressure, respectively, at spatial location $\textbf{x}$ at time $t$. $\textbf{F}({\textbf{x}},t)$ is the force per unit area applied to the fluid by the immersed boundary, i.e., the swimmer. These three quantities are modeled in an Eulerian framework on a fixed rectangular mesh. $\rho$ and $\mu$ are the fluid's density and dynamic viscosity, respectively.   

All \textit{interactions} between the swimmer and fluid are governed by integral equations with delta function kernels. As the swimmer bends, deformation forces are spread from its body to the nearest fluid mesh points. Similarly, the fluid velocity is interpolated back onto the swimmer to ensure the no-slip condition is satisfied. The integral equations that govern these dynamics are given as:
\begin{align}
   {\bf F}({\bf x},t) &= \int {\bf f}(s,t)  \delta\left({\bf x} - {\bf X}(s,t)\right) ds \label{eq:force1} \\
   \frac{\partial \textbf{X}}{\partial t}(s,t)  &= \int \textbf{u}({\bf x},t)  \delta\left({\bf x} - {\bf X}(s,t)\right) d{\bf x} \label{eq:force2}.
\end{align}
${\bf X}(s,t)$ and ${\bf f}(s,t)$ give the Cartesian coordinates and deformation forces along the immersed boundary (the swimmer) for each point denoted by Lagrangian parameter, $s$, and at time, $t$, respectively. Eqns (\ref{eq:force1})-(\ref{eq:force1}) essentially transform Lagrangian variables to Eulerian variables and vice versa. Here $\delta({\bf x})$ is a 2D delta function. These delta functions help ensure that forces from the immersed body are spread only to the nearest fluid mesh points to the immersed boundary, and vice-versa for when the velocity field is interpolated back to immersed boundary.

To construct a discretized 1D swimmer with physical meaning, throughout the entire swimmer's body linear springs and beams are used to connect adjacent Lagrangian points. The deformation force equations for springs and beams are given as the following,
\begin{align}
    \label{fiber_spring} \mathbf{F}_{spr} &= -k_{spr} \left( 1 - \frac{R_L}{\left|\left| \mathbf{X}_{F} - \mathbf{X}_L \right|\right| } \right) \cdot \left( \mathbf{X}_L - \mathbf{X}_F \right). \\
    \label{fiber_beam} \mathbf{F}_{beam} & =-k_{beam} \frac{\partial^4}{\partial s^4}\Big( \mathbf{X}(s,t) - \mathbf{X}_B(s,t) \Big),
\end{align}
where $k_{spr}$ and $k_{beam}$ are the spring stiffness and beam stiffness coefficients, respectively. In (\ref{fiber_spring}), the terms $X_{L}$ and $X_{F}$ represent the Cartesian positions of two Lagrangian nodes to which are connected by a spring - a leader (L) and a follower (F) node, at time, $t$. $R_L$ is that spring's corresponding resting length. In (\ref{fiber_beam}), $\mathbf{X}_B(s,t)$ represents the preferred curvature (shape) of the swimmer's body at time, $t$. This model dynamically changes $\mathbf{X}_B(s,t)$ over time, interpolating changing between preferred curvature states. Therefore this self-propelled swimmer propagates forward only due to time-varying body curvature and not through explicit prescribed motion of its Lagrangian points.

\texttt{IB2d} discretizes each beam using $3$ Lagrangian points; define those three points as
\begin{align}
\nonumber\textbf{X}(s,t) &= (X_q, Y_q), \\
\label{app:beam:lag}\textbf{X}(s+1,t) &= (X_r, Y_r), \\
\nonumber \textbf{X}(s-1,t) &= (X_p, Y_p). 
\end{align}

As described in \cite{BattistaIB2d:2018}, since Newton's Second Law of Motion relates forces to an overall acceleration, (\ref{fiber_beam}) is further discretized by a second derivative with respect to time. Thereby, all deformation forces arises from beams can be calculated in the following manner
\begin{align}
\nonumber {F}_{beam}(s-1,1) &= - k_{beam} \Bigg( \begin{array}{l}  X_r - 2X_q + X_p - C_x(t) \\   Y_r - 2Y_q + Y_p - C_y(t) \\ \end{array} \Bigg), \\
\label{app:beam:expForce} {F}_{beam}(s,1) &= 2 k_{beam}  \Bigg( \begin{array}{l} X_r - 2X_q + X_p - C_x(t) \\   Y_r - 2Y_q + Y_p - C_y(t) \\ \end{array} \Bigg), \\
\nonumber {F}_{beam}(s+1,1) &= -k_{beam} \Bigg( \begin{array}{l}  X_r - 2X_q + X_p - C_x(t) \\   Y_r - 2Y_q + Y_p - C_y(t) \\ \end{array} \Bigg).
\end{align}
Note that $C_x(t)$ and $C_y(t)$ are the preferred curvatures at time $t$. In these discretizations they are defined as
\begin{align}
\textbf{C} = \Bigg( \begin{array}{c} C_x \\ C_y \end{array} \Bigg) =  \Bigg( \begin{array}{c} X_{r_{pref}} - 2X_{q_{pref}} + X_{p_{pref}} \\ Y_{r_{pref}} - 2Y_{q_{pref}} + Y_{p_{pref}}  \end{array} \Bigg),
\end{align}
where the $pref$ subscript denotes the preferred geometric configuration. These quantities all may be time-dependent.

To discretize (\ref{eq:force1}) and (\ref{eq:force2}) regularized delta functions from \cite{Peskin:2002} were used , i.e., $\delta_h(\mathbf{x})$, 
\begin{equation}
\label{delta_h} \delta_h(\mathbf{x}) = \frac{1}{h^3} \phi\left(\frac{x}{h}\right) \phi\left(\frac{y}{h}\right) \phi\left(\frac{z}{h}\right) ,
\end{equation}
where $\phi(r)$ is defined as
\begin{equation}
\label{delta_phi} \phi(r) = \left\{ \begin{array}{l} \frac{1}{8}(3-2|r|+\sqrt{1+4|r|-4r^2} ), \ \ \ 0\leq |r| < 1 \\    
\frac{1}{8}(5-2|r|+\sqrt{-7+12|r|-4r^2}), 1\leq|r|<2 \\
0 \hspace{2.1in} 2\leq |r|.\\
\end{array}\right.
\end{equation}

\end{document}